\documentclass[iop]{emulateapj}
\submitted{Received December 31; published in ApJ July 1}

\usepackage{color}

\shorttitle{NGVS. IX. Abundance Matching in the Virgo Cluster}
\shortauthors{GROSSAUER ET. AL.}

\usepackage{etoolbox}
\usepackage{units}

\newcommand{\lcdm}{$\rm{\Lambda}$CDM}

\newcommand{\rvir}{\ensuremath{r_{\rm vir}}}

\newcommand{\minf}{\ensuremath{M_{\rm inf}}}
\newcommand{\mmax}{\ensuremath{M_{\rm max}}}
\newcommand{\mcur}{\ensuremath{M_{0}}}
\newcommand{\mhalo}{\ensuremath{M_h}}
\newcommand{\mstar}{\ensuremath{M_{\ast}}}

\newcommand{\vlos}{\ensuremath{v_{\rm los}}}
\newcommand{\zinf}{\ensuremath{z_{\rm inf}}}
\newcommand{\tinf}{\ensuremath{t_{\rm inf}}}

\newcommand{\vecsym}[1]{\ensuremath{\vec{#1}}}

\newcommand{\thead}[1]{\multicolumn{1}{c|}{#1}}
\newcommand{\theadfirst}[1]{\multicolumn{1}{|c|}{#1}}

\newtoggle{use-basic-table}
\toggletrue{use-basic-table}			

\iftoggle{use-basic-table}
{
	\newcommand{\zs}{\phantom{0}}
}{
	\newcommand{\zs}{}
}

\begin{document}
\title{The Next Generation Virgo Cluster Survey. IX. Estimating the Efficiency of Galaxy Formation on the Lowest-Mass Scales}
\author{Jonathan Grossauer\altaffilmark{1}, James E. Taylor\altaffilmark{1}, Laura Ferrarese\altaffilmark{2}, Lauren A. MacArthur\altaffilmark{2,3}, Patrick C{\^o}t{\'e}\altaffilmark{2}, 
Joel Roediger\altaffilmark{2}, St{\'e}phane Courteau\altaffilmark{4}, Jean-Charles Cuillandre\altaffilmark{5}, Pierre-Alain Duc\altaffilmark{5}, Patrick R. Durrell\altaffilmark{6}, 
S. D. J. Gwyn\altaffilmark{2}, Andr{\'e}s Jord{\'a}n\altaffilmark{7}, Simona Mei\altaffilmark{8,9}, Eric W. Peng\altaffilmark{10,11}}

\altaffiltext{1}{Department of Physics and Astronomy, University of Waterloo, Waterloo, ON, N2L 3G1, Canada; jgrossau@uwaterloo.ca, taylor@uwaterloo.ca}
\altaffiltext{2}{Herzberg Institute of Astrophysics, National Research Council of Canada, Victoria, BC, V9E 2E7, Canada}
\altaffiltext{3}{Department of Astrophysical Sciences, Princeton University, Princeton, NJ 08544, USA}
\altaffiltext{4}{Department of Physics, Engineering Physics and Astronomy, QueenÕs University, Kingston, ON, K7L 3N6, Canada}
\altaffiltext{5}{Laboratoire AIM Paris-Saclay, CNRS/INSU, Universit{\'e} Paris Diderot, CEA/IRFU/SAp, F-91191 Gif-sur-Yvette Cedex, France}
\altaffiltext{6}{Department of Physics and Astronomy, Youngstown State University, One University Plaza, Youngstown, OH 44555, USA}
\altaffiltext{7}{Departamento de Astronom\'ia y Astrof\'isica, Pontificia Universidad Cat\'olica de Chile, Av. Vicu\~na Mackenna 4860, Macul 7820436, Santiago, Chile}
\altaffiltext{8}{Universit{\'e} Paris Diderot,  F-75205 Paris Cedex 13, France}
\altaffiltext{9}{GEPI, Observatoire de Paris, Section de Meudon, 5 Place J. Janssen, F-92195 Meudon Cedex, France}
\altaffiltext{10}{Department of Astronomy, Peking University, Beijing 100871, China}
\altaffiltext{11}{Kavli Institute for Astronomy and Astrophysics, Peking University, Beijing 100871, China}

\begin{abstract}
The Next Generation Virgo Cluster Survey has recently determined the luminosity function of galaxies in the core of the Virgo cluster down to unprecedented magnitude and surface brightness limits. Comparing simulations of cluster formation to the derived central stellar mass function, we attempt to estimate the stellar-to-halo-mass ratio (SHMR) for dwarf galaxies, as it would have been before they fell into the cluster. This approach ignores several details and complications, e.g.,~the contribution of ongoing star formation to the { present-day} stellar mass of cluster members, and the effects of adiabatic contraction and/or violent feedback on the subhalo and cluster potentials. The final results are startlingly simple, however; we find that the trends in the SHMR determined previously for { bright galaxies} appear to extend down in a scale-invariant way to the faintest objects  detected in the survey. These results extend measurements of the formation efficiency of field galaxies by two decades in halo mass or five decades in stellar mass, down to some of the least massive dwarf galaxies known, with stellar masses of $\sim 10^5 M_\odot$. 

\end{abstract}
\keywords{dark matter -- galaxies: clusters: individual (Virgo) -- galaxies: dwarf -- galaxies: formation -- galaxies: general -- galaxies: luminosity function, mass function}

\section{Introduction}
\label{sec:introduction}

Galaxy formation is complicated; even a sketch description of the process involves cosmological structure formation in the dark matter component, gas cooling, disk formation, molecular cloud formation, star formation and stellar feedback from winds and supernovae, central black hole growth, feedback from an active nucleus, secular evolution, environmental effects, mergers, cold gas accretion, and the contribution from an intergalactic ionizing background { \citep[see e.g.,][for an overview and further references]{Motext}}. Each of these elements in turn involves complex, multi-scale physics that we are only starting to appreciate. Faced with this complexity, {\it ab initio} models of galaxy formation must necessarily be calibrated by comparing their global predictions with observations. Theoretical arguments and dissipationless simulations do provide a simple, robust  framework for understanding the abundance and clustering of cosmological structures, however -- the conventional picture of cold dark matter (CDM) halo formation {  \citep{FrenkWhite}}. How galaxies occupy CDM halos is then an empirical question that can be addressed directly by large surveys, without {  explicit} reference to the complex physics responsible for the final result. 

On intermediate mass scales, the relationship between galaxies and CDM halos  appears to be straightforward and one-to-one; a galaxy like the Milky Way includes within itself most of the stellar mass contained in its surrounding dark matter halo, and any halo as massive as that of the Milky Way probably hosts a similar dominant luminous galaxy. At the larger and smaller ends of the halo mass scale, however, the process of galaxy formation becomes more complicated and less efficient. The largest (galaxy cluster) halos contain thousands of individual galaxies rather than a single dominant one, and most of their baryonic mass is in {  the form of} hot gas, not stars. The smallest halos, on the other hand, do not appear to contain any galaxies or stars at all. It remains unclear exactly where in the mass spectrum the efficiency of galaxy formation drops to zero, and exactly which processes suppress star formation on which scales. Over some intermediate range of masses, however, we can take the ratio of the stellar mass of the central galaxy to the total mass of the surrounding halo as a simple indicator of the net efficiency of galaxy formation. This stellar-to-halo-mass ratio  \citep[SHMR -- e.g.,][and references therein]{Leauthaud12, Behroozi13, Moster13} has now been measured for a wide range of systems, either individually, using internal kinematics to estimate total mass {  \citep[e.g.,][]{Blanton08,Miller14}
or as an average in well-defined samples, using galaxy-galaxy lensing \citep[e.g.,][]{Leauthaud12, Velander14, Han15, Hudson15,Shan15}, 
satellite kinematics \citep[e.g.,][]{Conroy07,More11,Wojtak}, and overall abundance and/or clustering \citep[e.g.,][]{Behroozi13,Moster13,Rod15} to estimate the average halo mass for the sample.} The observationally derived SHMR provides a solid point of comparison for models of galaxy formation, bridging the gap between large statistical samples from surveys 
and detailed models or simulations of individual galaxies.
 
From the measurements to date, the SHMR has several interesting features. It reaches a maximum, corresponding to a peak in the efficiency of galaxy formation, at a characteristic halo mass comparable to the Milky Way's, $M_h \sim 10^{12} M_\odot$. At higher and lower masses, the ratio is roughly a power law (i.e.,~$\mstar/M_h \propto (M_h)^{\alpha}$, where the slope $\alpha$ has different values for small and large halo masses), although at either end of the scale it remains poorly constrained. Observations are beginning to probe the evolution of the ratio with redshift; the results are complex, with evidence for non-monotonic, mass-dependent evolution \citep{Behroozi13,Moster13}.
The SHMR is generally very hard to measure, particularly for low-mass galaxies; at low masses and high redshifts the measurement is essentially impossible, since we cannot measure dynamics for individual systems and cannot get enough signal from stacked samples of dwarf galaxies to determine an average halo mass from lensing. This is regrettable, given the suggestion that the SHMR changes significantly at the low-mass end as one goes to higher redshift \citep{Behroozi13, Miller14}. 

In the local universe, the smallest dwarf galaxies are highly clustered, occurring mainly as satellites of brighter systems \citep{Karachentsev13}. Even in the Local Group, where {  the dwarf population is best studied}, the satellites of the Milky Way and Andromeda seem very strongly clustered compared to the predicted distribution of halo substructure \citep{Kravtsov04,Taylor04}. In the context of structure formation, this biasing suggests that dwarfs are an old population, having formed independently and then been accreted into larger systems at high redshift. The best place to look for ancient, low-mass galaxies in the local universe should therefore be in the densest regions. The centers of the two main halos comprising the Local Group host the faintest known dwarf galaxies, but current samples are limited and incomplete; for larger and possibly older samples, we must search in the cores of nearby groups and clusters. The nearest large cluster, Virgo, contains nearly 900 galaxies brighter than $L_B = 3\times10^{7} L_{B,\odot}$ {  \citep{VCC}}, versus 10 in the Local Group \citep{McConnachierev}, and has a total mass of  $M_{\rm 200c, Virgo} = 5.2\times10^{14} M_\odot$ \citep{VirgoMassHigh, NGVS}, versus $M_{\rm 200c, LG} = 2.66\times10^{12} M_\odot$ \citep{vandermarel12}, so it represents a much richer hunting ground for low-mass structure.

The Next Generation Virgo Cluster Survey \cite[NGVS -- ][Paper I hereafter]{NGVS} is a wide-area, multi-band imaging survey of the Virgo cluster using MegaCam on the Canada--France--Hawaii Telescope (CFHT). It represents the first major revision to our picture of Virgo in the optical since the work of Binggeli, Sandage, and Tammann more than 25 years ago \citep{VCC}. Other papers in the NGVS series related to the topics considered here include those on the distribution of globular clusters in Virgo \citep{Durrell14}, the properties of star clusters, UCDs and galaxies in the cluster core (\citealt{Zhu,Zhang15}; Liu et al.~in preparation), the dynamical properties of low-mass galaxies (\citealt{Guerou15}, Toloba et al.~in preparation), interactions within possible infalling galaxies \citep{Paudel}, and optical-IR source classification \citep{Munoz14}.

Using newly developed techniques for flat-fielding and scattered light removal (Paper I; \citealt{Duc2015}; Cuillandre et al.~2015, in preparation),
the survey is extremely sensitive to low-mass and low-surface-brightness dwarfs. In principle, if we could relate the resulting stellar mass function to a predicted subhalo mass function, derived from a simulation of cluster formation, we could determine the SHMR at extremely low masses. Furthermore, since most of the galaxies in the core of Virgo were incorporated into this structure long ago, they should reflect the SHMR {  of field galaxies} at high redshift. 

There are a number of complications in carrying out this comparison, however. First, galaxies in clusters correspond to subhalos within larger halos. To determine the field SHMR we need to reconstruct the {\it original} mass a given subhalo had when it was first incorporated into the cluster. Second, baryons in the form of gas or stars may affect the dynamical evolution of subhalos, helping them survive the tidal mass loss and disruption seen in CDM-only simulations. Ongoing star formation will also add to the stellar mass of the galaxy, so the present-day stellar mass may not reflect its original mass on infall. Finally, obtaining a reliable estimate of the CDM substructure mass function in the core of a dense cluster is numerically challenging, since these regions are subject to the strongest resolution effects.

In this paper we generate a set of high-resolution simulations of CDM halos with properties similar to the Virgo cluster. 
From the simulations we determine the mean subhalo `infall' mass function (SIMF), that is the abundance of subhalos as a function of the mass they had at infall, for subhalos now at the center of the cluster, as well as the distribution of `infall redshifts' at which these objects were first incorporated into the larger structure. { (The SIMF corresponds to the `subhalo initial mass function' discussed in \citealt{TB04} or the `unevolved subhalo mass function' discussed in \citealt{vandenBosch05,Giocoli08,Giocoli10,Jiang14})}. We introduce several different models of the SIMF in an attempt to correct both for resolution effects and for the physical differences between subhalos and galaxies. Assuming a monotonic relationship between the predicted infall mass distribution and the observed stellar mass distribution, we calculate the SHMR for these systems when they first formed in the field. Our method makes a number of simplifying assumptions, some of which require further validation through more detailed simulations, but it provides an initial estimate of the efficiency of galaxy formation at very low mass and moderate redshift.

The outline of the paper is as follows. In section \ref{sec:observations} we summarize the NGVS and the determination of the stellar mass function for the cluster core. In section \ref{sec:simulations} we describe our cluster simulations. In section \ref{sec:catalogs} we discuss the subhalo catalogs, and present three different models for the infall mass distribution.  In section \ref{sec:shmr} we match the subhalo mass function to the observed stellar mass function, and discuss uncertainties in the resulting SHMR. We summarize our results in section \ref{sec:conclusion}. In an appendix, we also review subhalo properties, and discuss the trends in infall redshift  with radial separation or velocity offset from the cluster center. Throughout this paper we assume a WMAP-7 \citep{WMAP7} \lcdm{} cosmology with $\Omega_{b}=0.045$\footnote[1]{Note we assume this baryon density to generate a realistic initial power spectrum, although the subsequent numerical evolution treats all matter as collisionless.}, $\Omega_{c}=0.226$, $\Omega_{\Lambda}=0.729$, $h = 0.703$, $n=0.966$, and $\sigma_8=0.809$. 

\section{The NGVS stellar mass function}
\label{sec:observations}

\begin{figure}
\epsscale{1.10}
\plotone{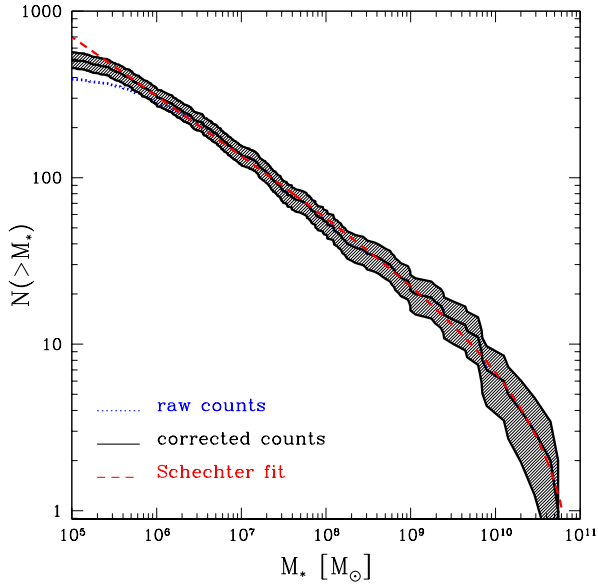}
\caption{The stellar mass function of Virgo galaxies in the cluster core (the `pilot region'), excluding M87. The dotted (blue) line indicates detected objects; the solid (black) line includes a correction for estimated incompleteness. The error contours indicate Poisson errors plus a fixed 10\%\ error accounting for systematics in the stellar mass determinations. The dashed (red) line is an integrated Schechter function fit with $M^* = 10^{12} M_\odot$, $\phi^* = 5.5$ and $\alpha = -1.35$.}\label{stellarmassfunction}
\end{figure}

The NGVS is a multi-band, panoramic survey of the Virgo cluster using MegaCam at the Megaprime focus on CFHT. An introduction to the survey, data analysis pipelines, data products and science goals is given in Paper I. Here we summarize information about the survey relevant to this work. 

The survey covers 104 square degrees around the two main components of the Virgo cluster, the A and B sub-clusters, extending out to 1--1.5 times the projected virial radius, depending on the estimated mass, in four bands {\it u*}, {\it g}, {\it i}, and {\it z} in the MegaCam filter system\footnote[2]{The survey was originally designed to include {\it r}-band coverage, although this is not yet complete outside the pilot region at the time of writing.} (for simplicity we will refer to these as $u,g,i$, and $z$ in the remainder of the paper). 
 
NGVS represents the first major update to our inventory of Virgo galaxies since the Virgo Cluster Catalog (VCC) of \cite{VCC}. The VCC covered most of the region surveyed by the NGVS, as well as part of the southern extension, with photographic plates in a single filter, reaching a depth of $B_{\rm lim}\sim 20$ mag for galaxies or 22--23 mag for point sources, and a surface brightness limit of  $\sim25$ mag\,arcsec$^{-2}$. By comparison, NGVS reaches a limit of 
$g= 25.9$ mag for point sources at S/N = 10, and a surface brightness limit of $29$ mag\,arcsec$^{-2}$ in {\it g}. The exceptionally faint surface brightness limit is the result of a new strategy for removing scattered light {  and treating large-scale spatial variations in the background}, {\it Elixir-LSB}, summarized in Paper I. 
The excellent seeing conditions at the CFHT site (NGVS images have a FWHM in the $ i$-band $ < 0.6^{\prime\prime}$) also allow us to resolve all but the most compact objects at the distance of Virgo, separating most ultra-compact dwarfs (UCDs) and even some of the more extended globular clusters from point sources. 

Given the timescale for obtaining, calibrating and analyzing survey data, the observations of an initial `pilot' region were analyzed first and it is these data we consider in this paper. We will present a detailed analysis of the full cluster stellar mass function in a subsequent paper. The pilot region consists of four contiguous fields (4 square degrees total area) around M87, the central galaxy of the main component of Virgo. These fields have complete coverage in all five bands, both in long exposures and in the short exposures designed to supplement our information in the saturated areas of the brightest galaxies. 

The identification and characterization of Virgo cluster members will be discussed in detail in an upcoming paper of this series (Ferrarese et al. 2015, in preparation); here we summarize the most relevant points. The detection of extended, low surface brightness Virgo galaxies in a field contaminated with a far larger number of foreground stars and background objects is a challenge -- conventional codes, such as SExtractor \citep{sextractor96}, would always regard low surface brightness objects as belonging to compact and/or brighter contaminants. To circumvent the problem, a ring median filter \citep{secker95} was first applied to each $g$-band image (the NGVS images with the highest signal-to-noise ratio), with radius adjusted to suppress all unresolved sources (stars and globular clusters) as well as compact background galaxies. A specific optimization of SExtractor was then run on the medianed {  stacks}, thus allowing the identification of low surface brightness sources that are potential Virgo candidates. The resulting object catalog does not of course discriminate between Virgo members and contaminants; therefore the next challenge is to define membership criteria. These are based on the location of each galaxy in a multi-parameter space defined by a combination of galaxy structural parameters (specifically size and surface brightness, measured using GalFit \citep{GalFit02}, in appropriately masked original images), photometric redshifts (based on $u,g,i,z$ and, when available, $r$-band photometry), and an index measuring the strength of residual structures in images created by subtracting from each galaxy the best-fitting GalFit model. The exact combination of axes in this space was selected to allow for maximum separation of known Virgo and background sources. The former comprise spectroscopically confirmed (mostly) VCC galaxies, while the latter are identified, using the same procedure described above, in four control fields located three virial radii away from M87 and presumed to be devoid of cluster members. 

The reliability of the procedure was tested in two independent ways. First, approximately 40,000 artificial galaxies were injected in the frames spanning the full range of luminosity and structural parameters expected for genuine Virgo galaxies, and then processed as described above. With few exceptions (for instance galaxies that land in the immediate vicinity of bright saturated stars or  near the cores of high surface brightness galaxies), all galaxies with surface brightness high enough to be visible in the frames were indeed recovered. A detailed discussion of the completeness of the data and the biases in the recovered parameters will be included in Ferrarese et al.~2015 (in prep.). Second, three NGVS team members independently inspected the 4 square degree of the pilot project region and identified, by eye, all objects that appeared to be bona-fide Virgo members; again with few exceptions, all such galaxies were detected and correctly identified as Virgo members by the code.

The photometry for each Virgo member was recovered using three separate techniques: by fitting a single Sersic \citep{sersic} component as implemented by GalFit; by fitting a core-Sersic, Sersic or double Sersic (for galaxies with a stellar nucleus) model to the 1-dimensional surface brightness profile derived by finding the elliptical isophotes that best fit the galaxy's surface brightness distribution; and by a non-parametric curve-of-growth analysis. There is generally excellent agreement between the magnitudes derived with the three different methods. All magnitudes were dereddened using reddening estimated at the location of the center of each galaxy following \cite{SFD}. 

Tests with the artificial galaxies also allowed for a strict quantitative assessment of our detection and completeness limits and galaxy parameter recovery. For a given magnitude, $g$, we describe the completeness, $f(g)$, as the fraction of input artificial galaxies recovered by the pipeline.  Scatter and bias in the recovered magnitudes were investigated and turn out to have only minor impact on the final stellar mass function, whereas completeness corrections become significant below $g=19$. Based on these tests, we fitted the recovered fraction as $f(g) = 1$ if $g < 18.9$ and $f(g) = 0.511 - 0.2 x+0.00444 x^3$ if $g > 18.9$, where $x\equiv g - 22$. The true number of objects was then estimated as $1/f(g)$ of those detected.

Stellar masses were obtained via Bayesian modeling of the $ugriz$ spectral energy distributions (SEDs) of the Virgo members. The integrated magnitudes of these galaxies were measured via a curve-of-growth analysis of the multi-wavelength imaging from the NGVS.  Errors were assigned to the photometry statistically by comparing the curve-of-growth magnitudes to those from an independent GALFIT analysis of the same imaging.  Our stellar population synthesis models span a multi-dimensional parameter space designed to mimic the wide variety of star formation histories (SFHs), chemical enrichment histories and dust contents of present-day galaxies.  We employed the base SSP models of \cite{BC03} and extinction was treated following the two-component prescription of \cite{Charlot00}.  A finite grid of ~50,000 synthetic stellar populations were then generated, assuming the priors described in \cite{daCunha}.  We fitted the SED of each galaxy using this grid to determine the marginalized posterior PDF for its stellar mass.  The final value corresponds to the median value of this PDF, with an uncertainty equal to half the interval between the 16th and 84th percentiles.

An important caveat regarding our modeling is the effect of bursts of SF on the resultant $\mstar/L$.  Our priors assume that bursts have a 50\%\ probability of occurring at each timestep throughout the lifetime of our synthetic populations, and that half of the models have not experienced a burst within the past 2 Gyr.  Since young (i.e.,~bright) stars have lower $\mstar/L$, it would be fair to say that our fiducial models are biased to low $\mstar/L$.  Reducing the contribution of bursts to the SFHs of our synthetic populations indicates that this bias could be as high as 0.2 dex.  However, given that the priors of our fiducial model have been tailored to reproduce the spectroscopic properties of SDSS galaxies, that the SFHs of real galaxies are unlikely to be smoothly varying, and that broadband colors cannot effectively constrain the role of bursts, we retain the $\mstar$ values predicted by our fiducial model for this analysis.  Further details on the SED modeling of NGVS galaxies will be presented in a forthcoming paper on the stellar {  populations} of the Virgo cluster (Roediger et al. 2015, in preparation). 

{  Finally, we note that we have not included UCDs in the core stellar mass function, since it is unclear whether to treat these as independent star-forming systems, or a population of unusually massive clusters associated with galaxies (see \citealt{Zhu,Zhang15}; Liu et al.~2015, in preparation, for a detailed discussion and references). The several hundred UCDs in the core region could potentially double our stellar mass function at the low-mass end, halving the halo masses derived for the smallest objects by abundance matching. Further work on the dynamics of these objects may clarify their cosmological status.} 

Figure \ref{stellarmassfunction} shows the cumulative stellar mass function for the core region, excluding M87. The dotted (blue) line indicates detected objects; the solid (black) line includes the correction for incompleteness. The error contours indicate Poisson errors plus a fixed 10\%\ error accounting for systematics in the stellar mass determinations. {  (Using SFHs with fewer bursts could further shift the stellar mass function up to 0.2 dec to the right, producing an increase in counts roughly equal to the 1-$\sigma$ error range shown here.)} Overall, for the purpose of this work, our sample consists of 407 galaxies with reliable stellar masses in the core region, spanning five decades or more in stellar mass. Below $\mstar \sim5\times 10^9 M_\odot$, the cumulative mass function is roughly a power law. An integrated Schechter function (ISF) $$N(> M) = \int_M^{\infty} \phi^* x^\alpha\exp(-M/M^*)dM/M^*,$$ with $M^* = 10^{12} M_\odot$, $\phi^* = 5.5$ and $\alpha = -1.35$ provides an excellent description of the stellar mass function (excluding M87) everywhere except at the smallest masses. Below $\mstar = 3\times10^5 M_\odot$, the mass function shows a possible change in slope, but the reality of this feature is unclear given systematic uncertainties in the completeness corrections at the faintest magnitudes. Thus in what follows we will adopt the ISF form with the parameters given above to represent the stellar mass function. 

The ISF has degeneracies in the model parameters, and we did not find that automated determination of all three parameters simultaneously worked particularly well. Instead, we determined experimentally that for fixed $M^* = 10^{12} M_\odot$, a plausible range of slopes is $\alpha = (-1.3)$--$(-1.42)$, and that the normalization can be in the range $\phi^* = 3$--8, depending on the value of $\alpha$. {  Allowing for SFHs with fewer bursts and systematically higher mass-to-light ratios broadens this range slightly, permitting a flatter fit with $\alpha \sim (-1.28)$ and $\phi^* = 12$}. In Figure \ref{Behroozi_comp2} below, we consider several extreme choices of {  ISF} parameter values that still provide a reasonable approximation to the data.

\section{Numerical Simulations}
\label{sec:simulations}

\begin{table*}
	\centering
	\iftoggle{use-basic-table}
	{
			\begin{tabular}{
		 | c
		 | c
 		 | c
		 | c
 		 | c
 		 | c
 	 	| c
 	 	| c
 	 	|}
	}{
		\sisetup{
		table-number-alignment = center,
		table	-figures-decimal  = 0,
		table-figures-integer  = 4
		}
		\begin{tabular}{
		 | S[table-figures-integer = 2 ,table-figures-decimal = 0]
		 | S[table-figures-integer = 1 ,table-figures-decimal = 2]
 		 | S[table-figures-integer = 2 ,table-figures-decimal = 1]
 		 | S[table-figures-integer = 4 ,table-figures-decimal = 0]
 		 | S[table-figures-integer = 4 ,table-figures-decimal = 0]
 		 | S[table-figures-integer = 4 ,table-figures-decimal = 0]
 		 | S[table-figures-integer = 2 ,table-figures-decimal = 1] 
	 	 | S[table-figures-integer = 1 ,table-figures-decimal = 2]
 		 |}
	}
		\hline
		\theadfirst{Halo} & \thead{ $\mhalo \left(10^{14} M_\odot \right)$ } & \thead{$N_{\rm part} \left(10^{6}\right)$} & \thead{$N_{\rm sub}$} & \thead{$\rvir \left({\rm kpc}\right)$} & \thead{$v_{\rm p}\left({\rm km s}^{-1}\right)$} & \thead{$c$} & \thead{$z_{50}$} \\
		\hline

		\zs{}3 & 3.67 &       20.9 & 5328 & 1868 &  \zs{}954 &  \zs{}4.1 & 0.42 \\
		\zs{}6 & 3.66 &       20.9 & 4983 & 1867 &  \zs{}986 &  \zs{}5.4 & 0.69 \\
		\zs{}7 & 2.04 &       11.7 & 2366 & 1536 &  \zs{}907 &  \zs{}9.8 & 1.30 \\
		      15 & 2.03 &       11.6 & 2546 & 1534 &  \zs{}827 &  \zs{}6.1 & 0.99 \\
		      20 & 2.39 &       13.7 & 3437 & 1619 &  \zs{}843 &  \zs{}4.9 & 0.67 \\
		      24 & 2.69 &       15.4 & 3656 & 1685 &  \zs{}883 &  \zs{}5.1 & 0.52 \\
		      27 & 1.54 &  \zs{}8.8 & 2154 & 1400 &  \zs{}841 &       10.6 & 0.39 \\
		      29 & 4.28 &       24.4 & 5786 & 1967 &        1071 &  \zs{}6.5 & 0.45 \\
		      35 & 1.86 &       10.6 & 2009 & 1489 &  \zs{}819  &  \zs{}6.9 & 0.66 \\
		      40 & 2.54 &       14.5 & 3756 & 1653 &  \zs{}862  &  \zs{}4.9 & 0.64 \\
		\hline
	\end{tabular}
	\caption{Properties of the resimulated halos: total halo mass $M_h$, number of particles $N_{\rm part}$, number of subhalos $N_{\rm sub}$, virial radius $r_{\rm vir}$, peak circular velocity $v_{\rm p}$, NFW concentration parameter $c$, half-mass accretion redshift $z_{50}$\label{tbl:resimSummary}}
\end{table*}
	
Abundance matching requires an estimate of the number of dark matter structures in a region, either from analytic theory or from numerical simulations of structure formation. In our case, the region in question corresponds to a line of sight through the core of a cluster, so most of the galaxies we detect should occupy subhalos within the larger cluster halo. While analytic and semi-analytic estimates of subhalo abundance do exist \citep[e.g.,][]{TB04,TB05a,TB05b,vandenBosch05,Giocoli08,Gan10,Yang11}, their accuracy is unclear in the densest environments. Thus, we will use numerical $N$-body simulations of halo formation to estimate the number of subhalos seen at small projected separation from the cluster core, and to determine their characteristic properties, e.g.,~their infall redshift, or their original mass at the time of infall. Our simulations include dark matter only, since the baryonic physics affecting the cluster mass distribution is complex and model-dependent. We discuss several specific baryonic processes which may affect our results in section \ref{sec:catalogs} .

In order to leverage the full power of the NGVS observations, we need simulations that resolve substructure down to very small masses. To accomplish this, we use the technique of resimulation, in which a high-resolution region of interest is embedded within a lower-resolution simulation. Multi-resolution initial conditions were generated using the \texttt{Grafic2} package \citep{Grafic2Paper}, and evolved using  the $N$-body code \texttt{Gadget2} \citep{GadgetPaper}. The initial, top-level simulation had $256^3$ particles in a cubic volume $140 \ h^{-1}$ Mpc on a side, large enough to contain many Virgo-mass clusters. With the chosen volume and WMAP-7 cosmology, this corresponds to a mass of $1.23\times 10^{10} \ h^{-1} M_\odot$ per particle. The softening length used was 0.02 times the mean inter-particle separation, or $\epsilon = 10.94 \ h^{-1}$ kpc in comoving units. In the top-level simulation, CDM halos were detected using the University of Washington friends-of-friends code FOF\footnote[3]{http://www-hpcc.astro.washington.edu/tools}. Candidate halos for resimulation were selected from the mass range 2--5 $\times 10^{14} M_\odot$, comparable to the estimated mass of the Virgo cluster (Paper I). In order to minimize interference from other large halos, any halo with a neighboring halo of greater than \nicefrac{1}{5} its mass within $3 \ h^{-1} $ Mpc was excluded from consideration. From the halos meeting these criteria, 10 were chosen at random for resimulation. These 10 halos represent a wide range of formation epochs and mass accretion histories.

Each of the selected halos was resimulated individually. For each halo, we determined the smallest rectangular volume in the initial conditions that contained all the particles in the final halo at $z=0$. This rectangular region was then extended to twice the linear size in each dimension, and all particles within this larger volume were replaced with higher-resolution initial conditions, using \texttt{Grafic2}. By choosing a larger volume around the cluster, we ensure that all particles associated with the final halo, out to roughly twice the virial radius, will be high-resolution particles, and that none of the more massive particles in the larger volume will intrude into the halo proper. 

The resimulations were performed with a factor of 1000 increase in mass resolution, the mass of the high-resolution particles being $1.23\times 10^7 \ h^{-1} M_\odot$. For the high-resolution particles, a softening length of 0.014 times the mean (high-resolution) inter-particle separation was used, $\epsilon_{\rm high} = 0.7656 \ h^{-1}$ kpc in comoving units. This choice corresponds to the optimal softening length for resolving substructure in halos of the selected mass range, as defined in \citet{SofteningPaper}. The softening length used for the low-resolution particles was the same as in the top-level simulation, $\epsilon_{\rm low}=10.9375 \ h^{-1}$ kpc. To record the assembly of each cluster we output 121 snapshots, equally spaced in $\log a$ between $z=0$ to $9$. We fit Navarro-Frenk-White \citep[NFW -- ][]{NFW97} profiles to the main cluster to determine its total mass, virial radius and concentration; neighboring halos and cluster subhalos were also detected in each snapshot, as described below. The structural properties of the resimulated halos are listed in Table \ref{tbl:resimSummary}. (Masses and virial radii assume the spherical collapse definition of the virial overdensity, rather than a fixed density contrast such as 200 times the critical density $\rho_{\rm c}$ or the matter density $\rho_{\rm m}$.) They contain from 8.8 to 24.4 million particles, equivalent to a mass of 1.54--4.28 $\times 10^{14} M_\odot$, and have several thousand resolved subhalos and sub-subhalos within their virial volume.

The clusters in our final sample vary significantly in mass, formation history and concentration, as indicated in Table \ref{tbl:resimSummary}. We could restrict the sample further, choosing a particular range of concentrations and/or formation histories based on more detailed dynamical or structural modeling of Virgo. Instead, we will retain the entire sample and use the cluster-to-cluster scatter as an indication of the systematic uncertainties in the subhalo mass function introduced by variations in the global structure and history of the cluster. We will normalize our results to a common cluster mass, however, since subhalo mass functions scale roughly with the mass of the parent halo. We take the mass of Virgo to be $M_{\rm 200,c}= 4.2 \times 10^{14} M_\odot$ and its concentration to be  $c=2.51$, following \cite{VirgoMassHigh}. Assuming an NFW profile with the corresponding scale radius, this  corresponds to a spherical collapse mass $M_{\rm vir}= 5.76 \times 10^{14} M_\odot$ and concentration $c_{\rm vir} \simeq 3.5$. These are the fiducial values adopted by the NGVS survey for the main component of Virgo, component A. Recent work by \cite{VirgoMassLow}, using X-ray spectroscopy and scaling relations, found a mass three times smaller for component A. It is also unclear whether component B, the other large component of Virgo, has merged with component A or should be considered a distinct halo. Component B has a mass of $M_{\rm 200,c}\sim 1\times 10^{14} M_\odot$ (Paper I), so this would increase the mass of the cluster by 20\%\ . Thus the mass of the cluster could conceivably lie in the range 0.33--1.2 times our fiducial value. We will discuss the implications of a different mass for Virgo in section \ref{sec:shmr}.

\section{Subhalo Catalogs}
\label{sec:catalogs}

From our resimulated halos, we then created subhalo catalogs to match to the NGVS observations. A rich variety of halo-finding techniques exist in the literature \citep[cf.][for a review]{AHFPaper}, but only a few are suited to finding subhalos within larger bound structures. To find subhalos in  the $z=0$ snapshot, we used the Amiga Halo Finder \citep[AHF --][]{AHFPaper}\footnote[4]{http://popia.ft.uam.es/AHF}. For the preceding snapshots, which were used only to identify the infall redshift and infall mass for each subhalo, we detected halos using the simpler and faster FOF code. We describe each of these steps below.

\subsection{The Subhalo Mass Function at $z = 0$}

\begin{figure}
\epsscale{1.10}
\plotone{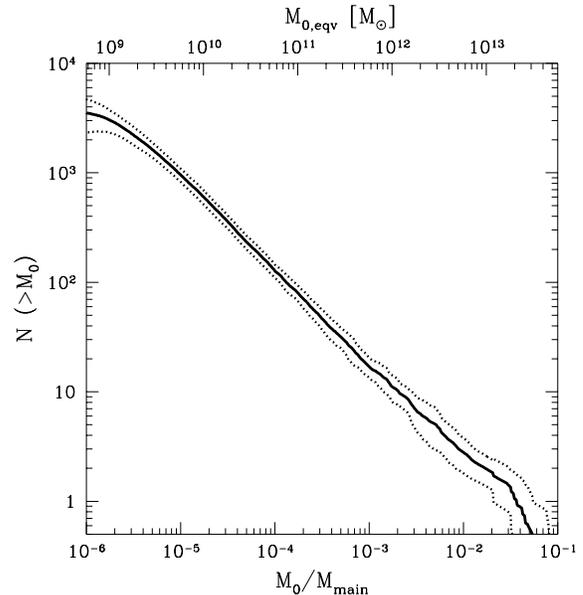}
\caption{The mean subhalo {  (self-bound)} mass function at redshift $z=0$, determined using the group finder AHF. Dotted lines show the 1-$\sigma$ simulation-to-simulation scatter. The lower axis shows subhalo mass at redshift zero, $\mcur$, relative to the virial (spherical collapse) mass of the main halo $M_{\rm main}$. The upper axis shows subhalo mass normalized to our fiducial (spherical collapse) mass for Virgo, $M_{\rm main} = M_{\rm vir}= 5.76 \times 10^{14} M_\odot$. \\ \label{fig:ahf_shmf}}
\end{figure}

In the $z=0$ snapshot, we need complete information about subhalos, including their position, velocity, final mass, and internal substructure. 
To derive the subhalo catalog we used the public halo finder AHF. AHF is a multi-scale, hierarchical group finder designed to run in parallel on large simulations, using the message-passing protocol (MPI). It constructs an initial representation of  the density distribution sampled on a coarse grid, identifies isolated regions that exceed some density threshold above the mean, and refines the grid iteratively in these regions. Structures are then identified as density peaks within this hierarchy, either at a base level in the field (halos) or as substructure (subhalos) within larger structures. As part of the process of associating particles with peaks, unbound particles are iteratively removed from each object, to ensure that it represents a bound physical structure. To identify halos at $z=0$, we use AHF's built-in calculation for the (spherical collapse) virial overdensity, rather than a fixed overdensity of 200 $\rho_{\rm c}$ or $\rho_{\rm m}$. Hereafter, we will refer to the basic subhalo catalog produced by AHF as `model 0'. 

Figure \ref{fig:ahf_shmf} shows the cumulative subhalo { (self-bound)} mass function obtained from AHF at $z=0$, using the AHF parameter value {\it NperRefCell} = 4, and a minimum group size of 20 particles. The thick solid line is the average for the sample of 10 clusters, while the dotted lines show the 1-$\sigma$ cluster-to-cluster scatter. The bottom axis shows subhalo mass at redshift zero, $\mcur$, relative to the mass of the main halo, while the top axis shows {  $M_{\rm 0,eqv}$, the subhalo} mass at redshift zero { rescaled} to our fiducial spherical collapse mass for Virgo, $M_{\rm vir} = 5.76 \times10^{14} M_\odot$. Consistent with many previous results \cite[e.g.,][]{Diemand07,Springel08,Gao11,Klypin11,Contini12,Gao12}, we find an approximate power-law with a logarithmic  slope $d\ln N(>M)/d\ln M \sim-0.9$ over almost four decades in mass, extending to a maximum mass ratio of $\mcur/M_{\rm main}\sim$2--5\%\ . The slope becomes slightly shallower around $M/M_{\rm main}\sim$ 2--3$\times 10^{-6}$ ($M_{\rm eqv}\sim10^9 M_\odot$), where resolution effects become important.

\subsection{Tracing Subhalo Histories}

The higher-redshift snapshots were used mainly to construct histories for each AHF subhalo detected at $z=0$, to determine when they last merged with a larger system. To construct the required merger trees, we used the simpler and faster group-finding code FOF\footnote[5]{See footnote 3.}, modified to work with multi-scale Gadget2 snapshots. The standard linking length of 0.2 times the mean inter-particle separation and a minimum group size of 20 particles were used to define the FOF groups.

{  We note there is a slight difference in our mass definitions in different time steps, since the AHF results at $z=0$ are spherical overdensity (SO) masses, whereas the results at higher redshift are FOF masses. \citet{Tinker08} compared the two types of mass estimates, and found that they agree to with 5-10\%\ on average, although SO masses can be up to 30\% smaller in cases where FOF artificially links nearby structures. AHF avoids these extended structures by using an additional unbinding step in its mass estimates. We note however that \citet{Tinker08} uses a SO method with an overdensity threshold of $200 \rho_m$, which produces masses up to 10\%\ higher than the AHF threshold at low redshift. As a result, our estimates of $M_{\rm inf}/M_{\rm main}$ may be 5-10\%\ larger than values calculated entirely using AHF, for systems merging at $z \lesssim 0.5$. Given that we are interested in broad trends over many orders of magnitude in mass, and that relatively few subhalos merge at redshifts this low, we will ignore this offset.}

To determine the merger history of each AHF subhalo, we proceed through the following steps:
\begin{enumerate}
\item First, we record all the subhalo's particles in the final snapshot, and search the preceding snapshot for structures (halos or subhalos) containing these same particles.  We define the {\it ancestor} of the subhalo as the structure containing the largest fraction of its particles in the preceding snapshot.
\item Iterating backwards through the snapshots from ancestor to ancestor, we determine an evolutionary history for each subhalo. We distinguish parts of the history where the ancestors are `merged' subhalos in a larger halo from parts where they are `independent' halos in the field. 
\item We define the maximum mass, $\mmax$, as the largest mass a subhalo's ancestor ever had as an independent halo. The infall redshift \zinf\ (or {  infall} time \tinf) is taken to be the last redshift (time) at which the ancestor existed as an independent halo before {  becoming part of} a larger system, and the infall mass $\minf$ is defined as the mass of the ancestor in the final snapshot before it  {  became part of} the larger system. 
{ \item At each step, the `main halo' is defined to be the {\it ancestor} of the main halo in the subsequent step, that is it is always the halo that contains the largest fraction of the particles of the main halo of the subsequent step. By tracing the main halo back step-by-step, we can define a `main trunk' of the merger tree, as opposed to the smaller `side branches' that merge with it.} 
\end{enumerate}

This analysis assumes that merging is a well-defined, one-way event. In fact,  FOF often links together, in the early phases of merging, structures which then separate back out into distinct halos in later steps. This `false merger' problem is discussed in detail in \cite{Fakhouri_Ma}. To reduce the influence of false mergers on our final statistics, we proceed forward in time through each subhalo's history and save the current state whenever its mass increases by more than \nicefrac{1}{3} between snapshots, indicating that a major merger has taken place. If the mass subsequently decreases by more than \nicefrac{1}{4}, this almost certainly represents the end of a false merger, so we revert to the last saved state and exclude {  from the calculation of the maximum mass} all ancestors between that snapshot and the snapshot under consideration. In this way, temporary and artificial jumps in mass from false mergers do not incorrectly reset the maximum mass.

It is not always possible to find $\minf$ and $\mmax$, especially for the smallest halos. It can be that a subhalo's ancestor was never independent, for instance, given our time resolution and starting redshift. This is not a significant problem, however, affecting less than 1\%\  of subhalos overall, and almost none of those with $\sim 200$ particles or more.

\subsection{Correcting for Completeness and Numerical Effects}

Three sorts of problems can arise when creating subhalo catalogs, that cause them to be incomplete. 
First, the group-finder may not correctly identify substructure in a given simulation output, or may not associate with a subhalo all the particles that are actually bound to it. This is especially likely for low-mass subhalos and/or subhalos in the dense core of the host halo \citep{Knebecomp}. Second, due to the discrete, $N$-body nature of the simulations, the dissipation of substructure is accelerated by numerical relaxation effects (e.g.,~\citealt{Moore_overmerging, Diemand_relax}; see \citealt{Moore_review} for a historical review of this problem). Thus the subhalos in the final snapshot will generally be less massive than they would have been if simulated at higher resolution, and some subhalos that should have survived to the present day will end up being completely disrupted. This problem affects the least massive and oldest  subhalos most strongly.  
Third, there is an important difference between the simulations and the physical process being modeled, namely the presence of baryons. Generically, baryons should cool and condense into the cores of subhalos, making them more resistant to mass loss and disruption. On the other hand, in some cases feedback from star formation or nuclear activity may eject mass from the center of the system, helping to disrupt the subhalo. Overall, we expect baryonic simulations to retain more substructure on galactic scales than the collisionless simulations considered here, but the details will be model-dependent \citep{Romano09,Romano10,Schew11}. 

While an alternative choice of group-finder might recover some of the missing structure \citep{Knebecomp}, numerical relaxation and the lack of baryons in the simulation are harder problems to address. Including baryons in the simulation brings with it the attendant problem of simulating galaxy formation correctly; and although overmerging can be decreased by increasing the resolution of the simulation, convergence is expected to be relatively slow, especially in the central regions \citep{Diemand_relax}. Instead, we will consider alternative approaches to recovering structure, based on subhalo merger histories. 

\subsection{Three Estimates of the Infall Mass Function}
\label{subsec:3models}

\subsubsection{Model 1 -- An Upper Limit on the Mass Function}
{ To overcome the issue of incompleteness with the subhalo catalogs, we can try a different approach. Every subhalo in the main halo was at one time an independent halo in the field. Thus, every subhalo will correspond to a merger event in the merger tree. We can therefore use these mergers as a way to identify subhalos. Similar `historical' approaches to finding substructure have been implemented previously with several group-finding codes \citep[e.g.,][]{Gill04,Tormen04}.
In the most conservative limit, we assume that every halo merging {\it at any point} in the merger tree, whether with the main trunk or with a side branch, survives  to the present day as a self-bound substructure. This approach is similar to those used by, e.g., \citet{Tormen04}, \citet{vandenBosch05}, and \citet{Giocoli08}, except that we include side branches, accounting for higher-order substructure (sub-subhalos), as in the semi-analytic models of \citet{TB04} or \citet{Giocoli10}.} In order to obtain $z=0$ subhalo properties, such as position and velocity, we further assume that the subhalo is traced by the most bound particle of its last independent ancestor. It is, of course, impossible to obtain $z=0$ subhalo masses with this method, but we can measure the mass at infall \minf\ for each subhalo. Assuming that every merger has a corresponding subhalo that survives down to $z=0$, we are certain to capture all actual subhalos down to some mass limit, so this approach (`model 1' hereafter) provides a conservative upper limit on the SIMF at $z=0$. Of course, model 1 also includes many systems that would not survive as distinct subhalos{ , since some will merge together physically or be disrupted, either before or after they enter the main halo. If, however,} we compare model 1 to the previous results from AHF (`model 0', defined above), which detects almost everything that does survive despite the factors that artificially accelerate the disruption of substructure {  in our simulations}, we can constrain the SIMF both { from} above and { from} below. 

\subsubsection{Model 2 -- An Intermediate Model}
Model 1 makes several unrealistic assumptions. First, it assumes that every system that merges with the main trunk of the merger tree represents a single, well-defined and dynamically relaxed halo. In the simulation, merging groups may have internal substructure or separate components that survive the merger as a distinct subhalos. Physically, these systems might correspond to individual galaxies in a loose group that are dissociated from one-another when they merge with the cluster. We can correct for this effect in part by identifying the least relaxed systems as they merge, and splitting them into their subcomponents. To do this we use the relaxation parameter $x_{\rm off} = {\left\vert\vecsym{x}_{\rm COM} - \vecsym{x}_{\rm MB}\right\vert}/r_{\rm vir}$, where $\vecsym{x}_{\rm COM}$ is the position of the group center of mass and $\vecsym{x}_{\rm MB}$ is the position of the most bound particle \citep{xoff}. If $x_{\rm off} > 0.25$, we consider the merging subhalo sufficiently unrelaxed that it will {  fall in} as one or more distinct components, and/or be tidally disrupted into these components by the present day. ($x_{\rm off} = 0.25$ is approximately the relaxation parameter expected for a 3:1 merger where the secondary has just crossed the virial radius of the primary.)

Second, we should also correct for the opposite limit, in which a sub-component of a subhalo remains bound {  to its parent subhalo} after infall, and has merged with its parent completely by the present day. We find this is mainly an issue for {{  massive subcomponents of systems merging} at $z \lesssim 1$. From dynamical friction arguments we expect merger or disruption rates to be fastest for sub-subhalos with a large mass relative to their parent. In particular, semi-analytic calculations \citep[e.g.,][]{TB04} suggest that the critical mass ratio for rapid disruption is around $M_{\rm sub}/M_{\rm main}\sim 0.1$. In practice, we consider that systems with $M_{\rm sub}/M_{\rm main} > 0.07$ will have merged if $t_{\rm inf}$ is less than half the age of the universe at $z=0$. The value 0.07 is chosen to match the model 0 results at the high-mass end of the subhalo mass function, where we expect the AHF substructure catalog to be complete and to provide good mass estimates.

Although these two corrections are approximate, they prune the model 1 results down to a more realistic set of subhalos. Applying the two corrections to the merger tree, we define a `model 2' for the infall mass function that resembles model 0 (the raw AHF results) at the high-mass end, and is intermediate between models 0 and 1 at the low-mass end.

\begin{figure}
\epsscale{1.10}
\plotone{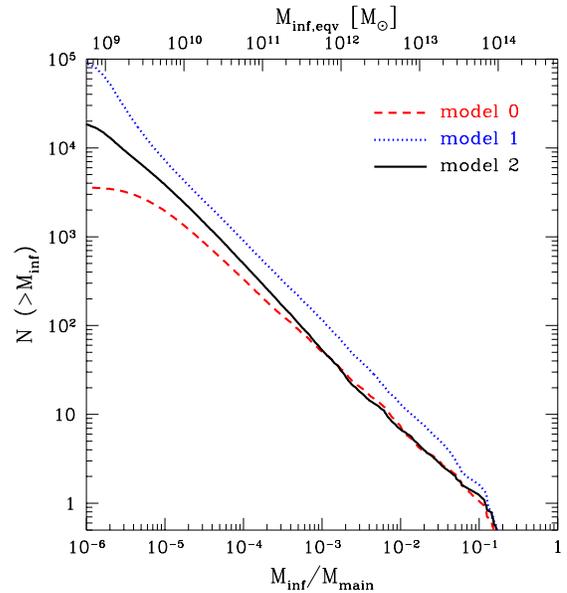}
\caption{The mean subhalo infall mass mass function (SIMF), using the three different models described in the text. The dashed (red) curve includes only the subhalos found by AHF (model 0). The dotted (blue) curve includes every object that ever merged into the merger tree
(model 1), while the solid (black) curve attempts to correct for loosely bound substructure and substructure disruption (model 2). The upper axis shows subhalo mass { rescaled} to the fiducial mass of Virgo, as in Figure \ref{fig:ahf_shmf}.\label{fig:model_shmf_compare}}
\end{figure}

\subsubsection{The Three Models Compared}
The three substructure models are compared in Figures \ref{fig:model_shmf_compare}, \ref{fig:mmax_zinf_dist}, and \ref{fig:model_pv_compare}. 
Figure \ref{fig:model_shmf_compare} compares the mean infall mass functions predicted by the three models. {  Roughly speaking, the predicted form of the SIMF is a power-law with an exponential cutoff at $M_{\rm inf}/M_{\rm main} \sim 0.1$. The power-law slope $\alpha = d\ln N/d\ln M$ is $\sim -0.8$ for model 0, $\sim -0.96$ for model 1, and $\sim -0.87$ for model 2. These values are consistent with previous numerical and semi-analytic estimates of the SIMF, which is expected to be fairly universal \citep{vandenBosch05,Giocoli08}.  Methods that only count `first-order' (i.e.,~direct) branchings off the main trunk of the merger tree find slopes of $\sim -0.8$ or shallower \citep[e.g.,][]{TB04,Giocoli08}, while methods that account for higher-order branchings (or sub-subhalos) find slopes of $\sim -0.9$ \citep[e.g.,][]{TB04,Giocoli10,Jiang14}. Although it was not chosen to match these estimates, model 2 is quite close to the slope and normalization derived in the most recent semi-analytic work \citep{Jiang14}. 

Overall, the range in uncertainty between the three models} is roughly a factor of 2--3 except at the smallest masses, where numerical resolution renders the raw AHF results (model 0 -- lowest curve) increasingly incomplete. Model 2 predicts results similar to model 0 at large masses, but finds a factor of 2 or more subhalos at small masses. Figure \ref{fig:mmax_zinf_dist} shows subhalo abundance as a function of infall mass and infall redshift. We see that both models 1 and 2 (right and center panels) include a large number of early and/or low-mass subhalos, relative to model 0. On the other hand, all three models find similar numbers of massive, recently merged halos.

\begin{figure*}
\epsscale{1.00}
\plotone{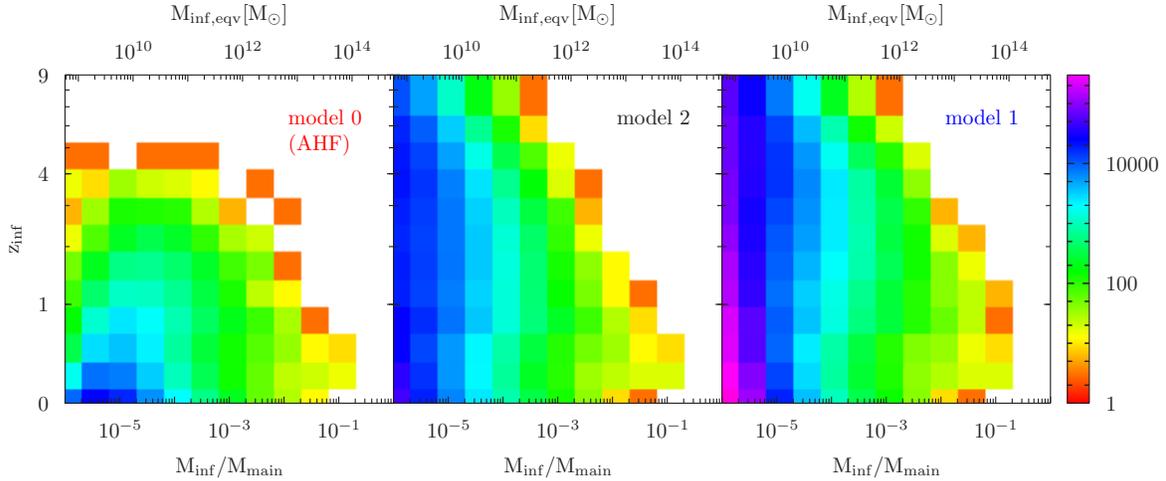}
\caption{Subhalo abundance as a function of  infall mass \minf\ and infall redshift $z_{\rm inf}$. The left and right-most panels show the results for model 0 (AHF) and model 1, respectively. { The central panel shows the results for the intermediate model 2.}\label{fig:mmax_zinf_dist}}
\end{figure*}

\begin{figure*}
\begin{center}
\epsscale{1.0}
\includegraphics[scale = 0.6,trim= 0 0 0 250, clip = true]{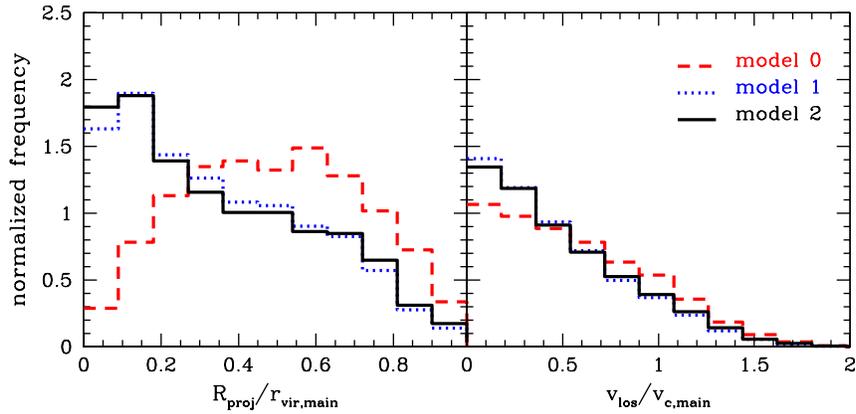}
\caption{The normalized distributions of projected radial separation from the cluster center (left panel) and velocity offset (right panel) for the three substructure models; colors and line-styles are as in Figure \ref{fig:model_shmf_compare}. \label{fig:model_pv_compare}}
\end{center}
\end{figure*}

Since older, more stripped subhalos will generally lie on more tightly bound orbits, we also expect the three models to predict different radial and velocity distributions. Figure \ref{fig:model_pv_compare} shows subhalo abundance as a function of radial offset from the cluster center (left panel), and line of sight velocity offset (right panel), normalized to the virial radius and the circular velocity at the virial radius, respectively. There is a dramatic difference in clustering between model 0 (the raw AHF results) and models 1 and 2, but relatively little difference in clustering between models 1 and 2.
 
Thus, in summary, averaged over the whole virial volume model 1 predicts roughly twice the abundance of substructure at a fixed infall mass, while model 2 predicts numbers closer to model 0, but models 1 and 2 both have many more old, low-mass, and/or centrally located subhalos. When we restrict the subhalo sample to the central region of the cluster, the result is that the three models predict significantly different subhalo mass functions in this region. 

To compare the simulation results to the stellar mass function in the pilot region, we define an analogous region in each simulation: a beam with a square cross-section 0.252 $r_{\rm vir}$ on a side, passing through the center of the cluster along the line of sight. For our fiducial Virgo mass, this corresponds to the part of the cluster covered by the pilot region{ , as discussed in section \ref{sec:shmr} below}. Figure \ref{fig:centre-shmf} shows the SIMF for halos in this central region. Within the central region the mass functions for models 1 and 2 now differ by a factor of more than 5, highlighting the considerable uncertainties in subhalo abundance matching that are associated with the details of subhalo evolution. We will discuss the effect of these differences on abundance matching below; we also discuss the spatial and orbital properties of the subhalos further in appendix \ref{sec:subhalos}.

\begin{figure}
\epsscale{1.10}
\plotone{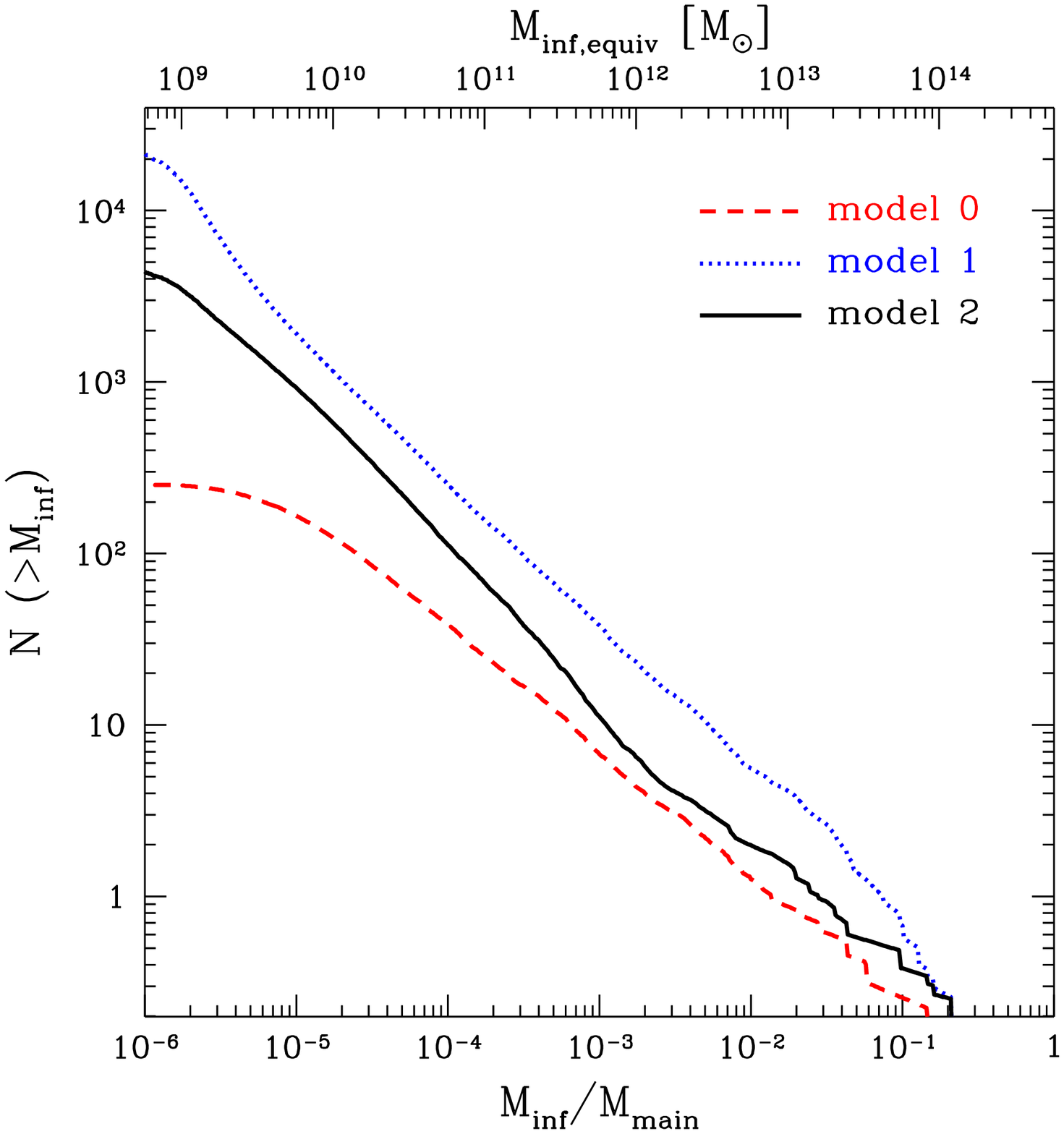}
\caption{The average SIMF of the central region for the three substructure models. As in Figure 3, the dashed (red) curve is for model 0, the dotted (blue) curve is for model 1, and the solid (black) curve is for model 2. \label{fig:centre-shmf}}
\end{figure}

\section{The Stellar-to-Halo Mass Ratio}
\label{sec:shmr}

\subsection{Basic Results in the Cluster Core}
To determine the subhalo SHMR, we use the technique of abundance matching, assuming the most massive satellite galaxy is hosted by the most massive subhalo, the next most massive by the next most massive subhalo, and so on. There are three definitions of the subhalo mass that we can use to construct this relationship: $\mcur$, the mass of the subhalo at $z=0$; $\minf$, the final mass the subhalo had as an independent halo, immediately prior to merging into a larger system; and $\mmax$, the largest mass the subhalo ever had as an independent halo. In practice, we find the results are almost identical for the latter two choices. Using $\mcur$ produces very different results (systematically lower halo masses for a given stellar mass), but this choice seems unphysical since tidal stripping may have reduced the dark-matter mass of a subhalo significantly without affecting its inner stellar component. Thus we consider \minf\ as the most logical choice for subhalo abundance matching. In particular, if the SHMR { were} independent of redshift, then we {  would} expect it to be comparable in the clusters and in the field, when expressed in terms of \minf.  {  (There is in fact evidence that the field SHMR varies slightly with redshift, as discussed below.)}

The fiducial model of the Virgo cluster adopted by NGVS has a total mass $M_{\rm 200,c}= 4.2\times 10^{14} M_\odot$ and a concentration $c=2.51$ with respect to the outer radius $r_{\rm 200,c} = 1.55$ Mpc, or  $5.38^{\circ}$ at the fiducial distance of $16.5$ Mpc. Assuming a NFW profile, the corresponding spherical collapse mass is 
$M_{\rm vir}= 5.76 \times 10^{14} M_\odot$ and the spherical collapse virial radius is $r_{\rm vir} = 2.16$ Mpc, or $7.5^{\circ}$ at the distance of Virgo, giving a spherical collapse concentration $c_{\rm vir} \sim 3.5$. The pilot region corresponds to a square patch 2$^\circ$ on a side (4 MegaCam pointings), centered on M87, the center of component A. Taking into account chip boundaries and edge effects, the effective area is 1.9$^\circ$ on a side, or $0.252\,\rvir$ on a side for our fiducial mass and distance.
Thus, in each of {  three orthogonal} projections of the resimulated halos, we select only the subhalos in a projected region of this size for comparison with the core stellar mass function, scaling the size of the region to the virial radius of each individual cluster. As explained in section \ref{sec:simulations}, we also rescale the mass of each cluster halo to our fiducial mass for Virgo, and adjust all the subhalo masses accordingly. The  30 sets of subhalos { (that is three projections of each of 10 clusters)} thus defined 
for each of the three substructure models comprise our simulation data for all of the following analysis, unless otherwise stated. We match these to the previously determined observed stellar mass function (see section \ref{sec:observations}), excluding the central galaxy M87, since it should be matched to the main halo, rather than any individual subhalo within it.

\begin{figure}
\begin{center}
\includegraphics[scale = 0.75,trim= 20 20 0 0, clip = true]{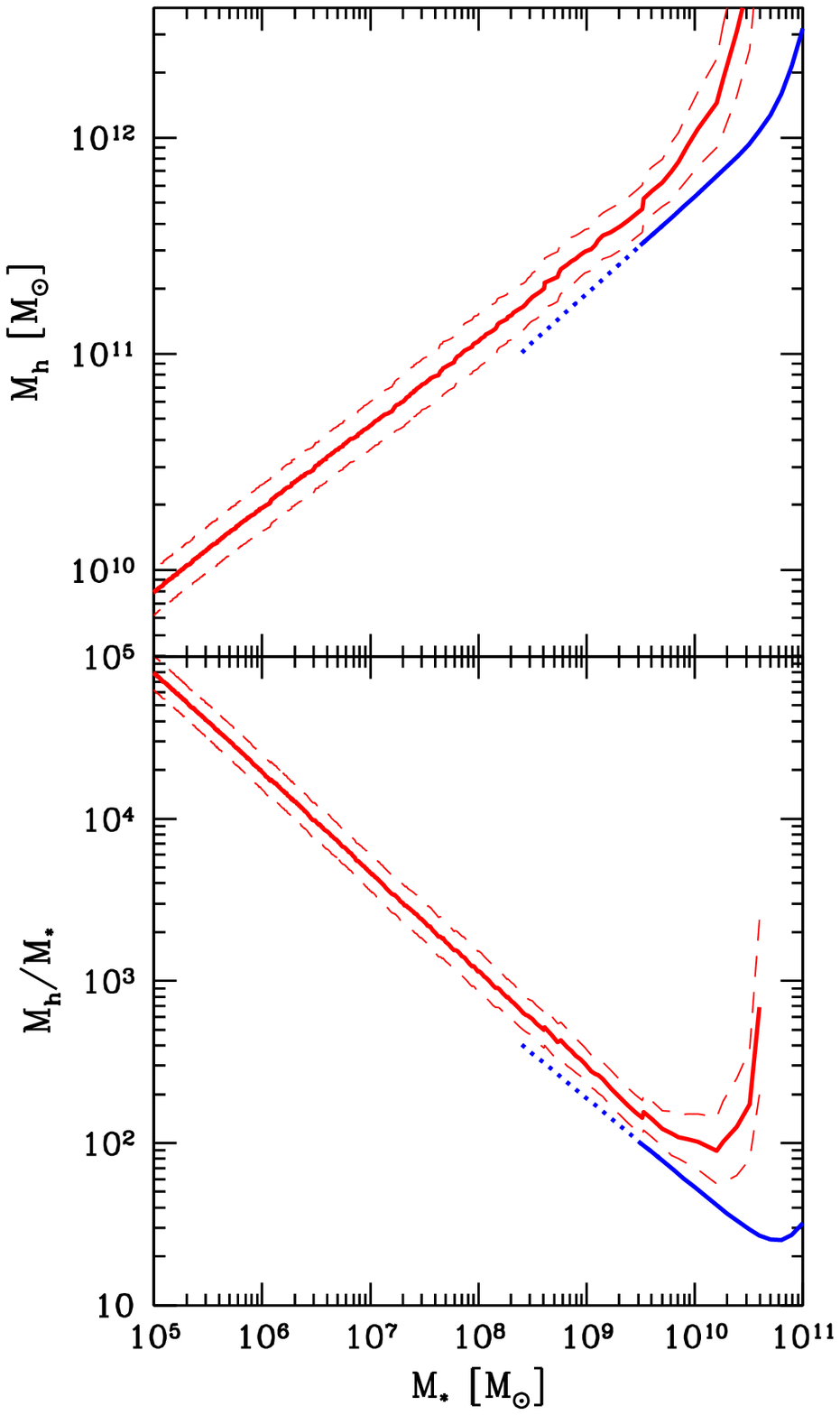}
\caption{Halo mass \mhalo  (top panel), and the ratio of halo to stellar mass { $M_h/\mstar$} (lower panel), as a function of stellar mass \mstar. The thin lines show the 1-$\sigma$ scatter of the simulations. The smooth blue curves at large stellar mass show the results from \cite{Leauthaud12} at  $z=0.88$ for comparison (solid, with an extrapolation of their fit to lower masses dotted).\label{fig:abdcmatch}}
\end{center}
\end{figure}

The average SIMF for the central region, for the three different substructure models, was shown in Figure \ref{fig:centre-shmf}. Using the results for the intermediate model 2, which represents our best guess at the true mass function, we match the SIMF {  for each of the 30 subhalo catalogs} to the stellar mass function of the cluster core { shown previously in Figure \ref{stellarmassfunction}}. The results are shown in Figure \ref{fig:abdcmatch}, as total (infall) halo mass versus stellar mass (top panel), or halo-to-stellar-mass ratio versus stellar mass (bottom panel) . The thick solid (red) line shows the mean relationship, while the thin dotted lines show the 1-$\sigma$ scatter amongst the 30 subhalo catalogs. The smooth curve at high stellar mass shows the results from the analysis of \citet[][-- L12 hereafter]{Leauthaud12}, which combines constraints from galaxy-galaxy lensing, galaxy clustering, and the stellar mass function, {  for comparison}\footnote[6]{Leauthaud et al. define halo mass using a density contrast of 200 times the matter density $\rho_{\rm m}$, so their halo masses will be a few percent larger than ours for identical halos at $z = 0.88$.}. 

{  We note that the results of Leauthaud et al.~are based on isolated field galaxies, whereas we are considering a population of galaxies now for the most part located deep in the core of a galaxy cluster. On the other hand, we are using the subhalo infall mass \minf, rather than the present-day mass \mcur, for comparison. If the present-day cluster members were typical field galaxies at the moment they fell in to the cluster, and if their stellar mass remained unchanged subsequently, we would expect their stellar-to-infall-mass ratio to match the SHMR for field galaxies at their infall redshift. Galaxies that fall into a cluster may not be completely typical, of course, as the conventional picture of structure formation predicts that they will always have resided in regions denser then the mean, even before they merged with the cluster. Comparing our stellar-to-infall-mass ratio to the field SHMR allows us to explore this possible environmental dependence. } The mean redshift for the {  L12 field} sample shown here is $\langle z\rangle$ = 0.88, which should be {  slightly lower than} the mean infall redshift of the subhalos, { $\langle z_{\rm inf}\rangle$ = 1.5 for model 2}. We also note that L12 only measured the SHMR down to $\log(M_h) \sim 11.5$ at this redshift; the dotted line shows an extrapolation of their power-law fit to lower masses, to aid comparison.

We see two main features in our results. First, at intermediate stellar masses our results agree roughly with those of Leauthaud et al., both in slope and in normalization. There may be a slight ($\sim$ 20\%) offset in normalization, but it is only marginally significant given the cluster-to-cluster scatter and the difference in redshift of the two samples, {  and is also well within the range of possible bias in our stellar mass-to-light ratios discussed previously in section \ref{sec:observations}}. Physical processes such as star formation or stripping of stellar material after infall could also affect stellar masses {  of cluster galaxies} relative to the field; we would need better statistical evidence to establish whether this is the case. 

Combining our results at low stellar masses with those of L12 at high stellar masses, it appears the SHMR can be fit by a single power-law $M_h \sim \mstar^{0.39}$ from masses of \mstar $\ll10^{6} M_\odot$ to \mstar $\gtrsim10^{10} M_\odot$ (although at that point the slope of the relationship starts to change). {  \cite{Behroozi10} have proposed a more precise functional form for the relation between halo and stellar mass, also used in L12:
\begin{displaymath}
\log_{10}(M_h) = \hspace{0.65\columnwidth}
 \end{displaymath}
 \vspace{-3ex}
$$\quad \log_{10}(M_1) + \beta\,\log_{10}\left(\frac{M_\ast}{M_{\ast,0}}\right) +
 \frac{\left(\frac{M_\ast}{M_{\ast,0}}\right)^\delta}{1 + \left(\frac{M_{\ast}}{M_{\ast,0}}\right)^{-\gamma}} - \frac{1}{2},\hspace{0.05\columnwidth}$$
Over the range $\log(\mstar/M_\odot)$ = 5.0 -- 10.5, our results are well-described by this form with parameters [$\log(M_1/M_\odot)$, $\log(M_{\ast,0}/M_\odot)$, $\beta$, $\delta$, and $\gamma$] = [12.45, 10.35, 0.39, 0.4, 1.0]. Given the uncertainties at the high-mass end, however, our data only really constrain the faint-end slope $\beta$ and the normalization (set by $M_1$ and $M_{\ast,0}$).} {\it Overall, measuring the abundance of faint dwarf galaxies in Virgo allows us to extend previous measurements of the SHMR several orders of magnitude in stellar mass, down to the faintest galaxies known outside the Local Group.} 

The second feature is a possible offset between our results and those of L12 at large stellar masses. For the few largest objects in the core of Virgo, our results suggest infall halo masses $\sim$2--3 times larger than the results of L12 at $\left<z\right> = 0.88$. This could be because the largest systems merged at higher redshift, when the SHMR was lower in this mass range (see Figure \ref{Behroozi_comp2} below), or it could indicate star formation has been suppressed in these systems, relative to the field. {  Bias in our SHFs could also shift our stellar mass scale up (i.e.,~to the right in Figure \ref{fig:abdcmatch}) by as much as to 0.2 dex, as discussed in section \ref{sec:observations}}. The statistics are poor at this end of the relationship, however, due to the small number of bright galaxies in the pilot region. Results from the full NGVS survey will allow us to explore this trend over the full range in stellar mass.

\subsection{Comparison to Recent Work}

\citet[][-- B13 hereafter]{Behroozi13} recently modeled  average galaxy star formation histories and estimated the SHMR for field galaxies over a broad range of mass and redshift. Assuming our results within the core of Virgo reflect the field SHMR at the infall redshift, they complement those of B13, covering most of the same redshift range, but extending down one to two orders of magnitude in halo mass at intermediate redshift. 

Figure \ref{Behroozi_comp} compares our derived SHMR to the results of B13 and L12. The thin colored lines show the models of Behroozi et al.~for $z=0.1$, 1, 2, 3, 4 and 5 (black, blue, cyan, green, yellow, and orange respectively), while the dashed lines show the results of Leauthaud et al.~for $z = 0.37$ and 0.88 (black and blue lines respectively). The thick black line shows our derived SHMR, and the dotted lines show the 1-$\sigma$ cluster-to-cluster scatter. {  Once again, our results lie slightly above the field SHMR at z=0--1, but are in fairly good agreement with the field estimates at redshift $z\sim$2, for the mass range $10^9 M_\odot$-- $10^{10} M_\odot$. There is a possible discrepancy for $\mstar \gtrsim 3\times 10^{10} M_\odot$, where we predict a factor of $\sim 2$ deficit of stellar mass associated with the most massive halos that merged into Virgo. Some or all of this offset could be related to the possible bias in our mass-to-light ratios discussed in section \ref{sec:observations}.} 

\begin{figure*}
\begin{center}
\includegraphics[scale = 0.8,trim= 0 0 0 150, clip = true]{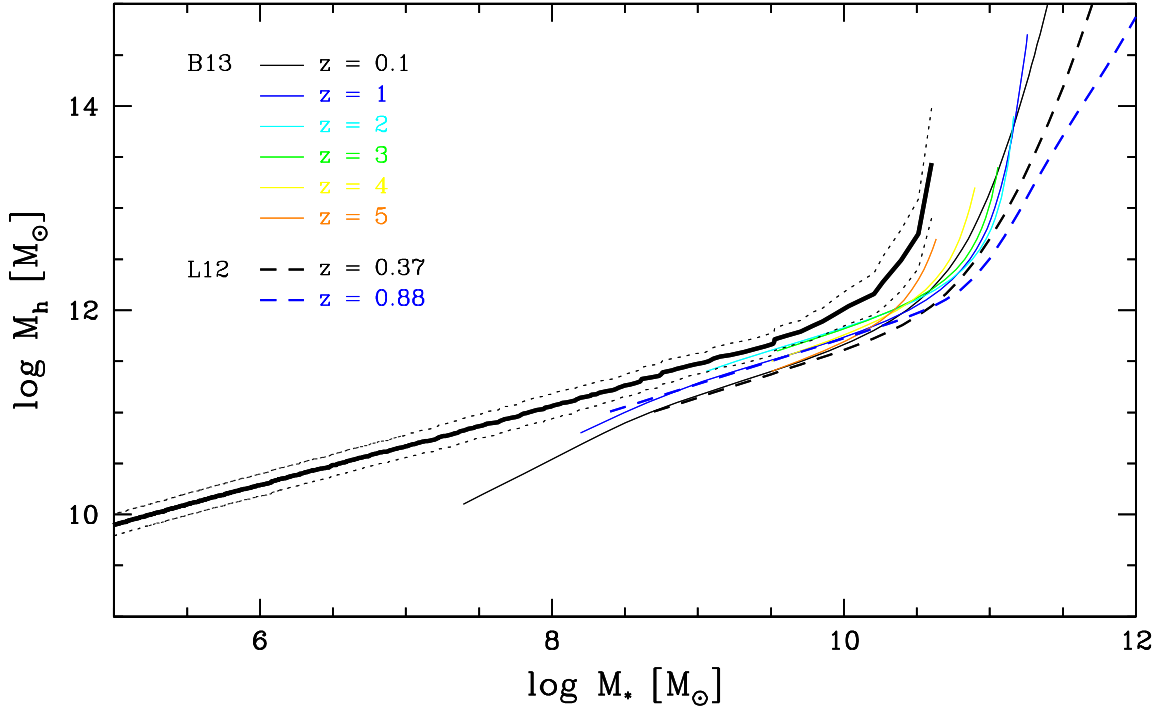}
\caption{The derived relationship between stellar mass and halo mass, compared to the results of B13 and L12. The thin colored lines show the models of Behroozi et al.~for $z=0.1$, 1, 2, 3, 4 and 5 (black, blue, cyan, green, yellow, and orange respectively), while the dashed lines show the results of Leauthaud et al.~for $z = 0.37$ and 0.88 (black and blue lines respectively). The thick black line shows our derived SHMR, and the dotted lines show the 1-$\sigma$ cluster-to-cluster scatter.}
\label{Behroozi_comp}
\end{center}
\end{figure*}

These preliminary results seem promising, but are subject to a number of uncertainties. Figure \ref{Behroozi_comp2} illustrates  a few of the main uncertainties in the modeling. The top left panel compares our mean SHMR as a function of halo mass (solid black line with dotted lines indicating the 1-$\sigma$ cluster-to-cluster scatter) to those of L12 and B13. As before, at large {  halo} masses, we predict less stellar mass {  for} a given halo mass. The discrepancy is particularly noticeable around the peak of star formation efficiency, $M_h \sim 10^{12}$, where the field SHMR is almost 3 times what we find in Virgo. Our results here are based on the small number (6-7 in total) of very massive galaxies in the cluster core, so shot noise introduces considerable uncertainty in the abundance matching. Nonetheless, our results suggest a lower star formation efficiency for the most massive galaxies in the core of Virgo. In the conventional picture of hierarchical structure formation, the progenitors of these galaxies have always occupied high-density regions, so it is plausible that their star formation may have been suppressed even before infall into {  the cluster}.

\begin{figure*}
\begin{center}
\includegraphics[scale = 0.8,trim= 0 170 0 0, clip = true]{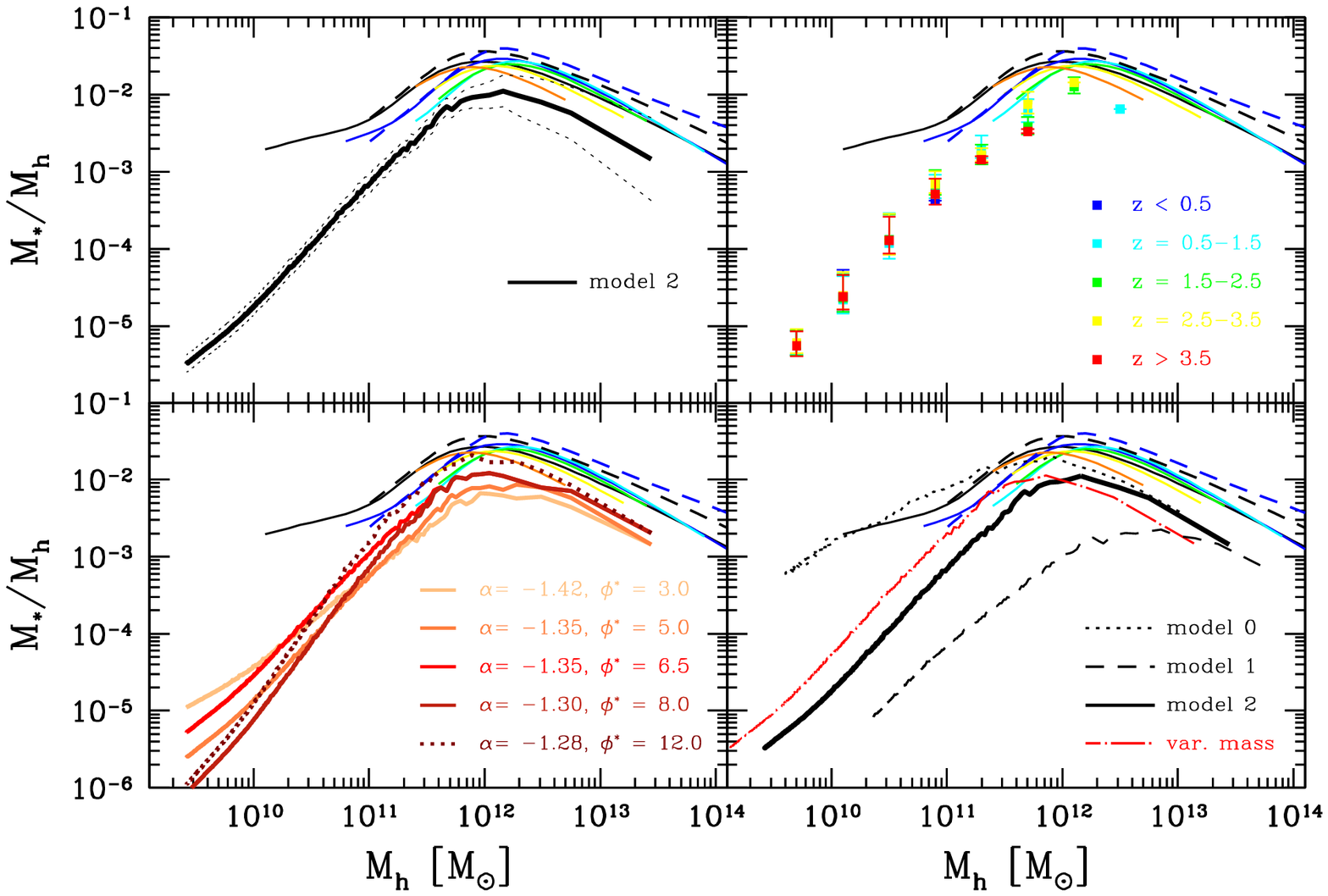}
\caption{The derived SHMR, compared to the results of B13 and L12. In each panel, the thin solid and dashed lines show the results of B13 and L12, with colors and line-styles as in Figure \ref{Behroozi_comp}. The top left panel shows the mean SHMR derived for model 2, with 1-$\sigma$ simulation-to-simulation scatter (thick black line and dotted black lines respectively). The top right panel shows individual mass ratios binned by infall redshift (points with error bars; colored as indicated). The bottom left panel shows the effect of alternate fits to the stellar mass function {  (solid lines), as well as the effect of fewer bursts in the SFHs (dotted line)}. The bottom right panel shows the SHMR derived for the two other subhalo counting models, model 0 (the raw AHF results -- upper dotted curve) and model 1 (lower dashed curve),  as well as the effect of reducing the mass of Virgo to \nicefrac{1}{3} of the fiducial value (dot-dashed curve).}
\label{Behroozi_comp2}
\end{center}
\end{figure*}

At lower halo masses, our results extend the trend seen between  $\sim 2\times 10^{11}$ and $10^{12} M_\odot$ down to halo masses of $\sim 5\times 10^{9} M_\odot$. This appears to conflict with the results of B13 at $z=0.1$ and $z=1$ (thin black and blue curves), but only a fraction of the subhalos in the core of the cluster have infall redshifts this low, as is apparent from Figures \ref{fig:mmax_zinf_dist}, \ref{fig:observables}, or \ref{fig:velocitybins}. Instead, our mean results are most sensitive to subhalos with $z \gtrsim 1$--2. 

Since our abundance matching is based only on $\minf$, independent of $\zinf$, we do not expect any significant redshift dependence in the final results. This is confirmed in the top right panel, where we show results for individual subhalos binned by infall redshift (points with error bars; colors correspond to $z < 0.5$ (blue), $0.5 \leq z < 1.5$ (cyan), $1.5\leq z<2.5$ (green), $2.5 \leq z < 3.5$ (yellow), and $z \geq 3.5$ (red)). Our derived SHMR is independent of $\zinf$ to within the scatter, as expected. The models of B13 and L12 predict some evolution in the SHMR, particularly at low masses, so this is something we hope to test for with the results of the full NGVS stellar mass function, as discussed further in appendix \ref{sec:subhalos}.

Uncertainties in the stellar mass function also affect the derived SHMR, as illustrated in the lower left panel. The four thick (solid) curves show a range of Schechter-function slopes and normalizations consistent with the pilot region stellar mass function, once photometric errors and population synthesis uncertainties are taken into account. Models with shallower slopes and/or higher normalizations are in excellent agreement with the results of B13 at $z=$0.5--2.5, {  for halo masses up to $\sim5\times10^{11} M_\odot$, although they still predict a lower SHMR at the largest halo masses. The dotted curve shows the effect of systematically higher mass-to-light ratios, if we have overestimated the importance of bursts in our SFHs, as discussed in section \ref{sec:observations}. This would eliminate the discrepancy with the field results, except for the few most massive systems (halo masses of $\sim 1$--2$\times 10^{12} M_\odot$ or more), where the cluster-to-cluster scatter is large.}

One shortcoming of our method is the systematic uncertainty associated with subhalo counting. The bottom right panel shows how the three models presented in section \ref{subsec:3models} predict dramatically different SHMRs. The raw AHF results (model 0) find relatively few surviving subhalos in the cluster core, so galaxies are matched to objects further down the mass function, producing an extremely high SHMR (upper dotted black line). This seems inconsistent with the field results of B13 at all redshifts, confirming that the raw AHF catalog probably underestimates subhalo abundance. The most conservative subhalo counting model, model 1 (lower dashed curve), predicts a SHMR 5--10 times lower than that of B13 at all stellar masses, which also seems implausible given the field results. Our intermediate model, model 2 (middle, solid curve) matches the results of B13 fairly well where the two overlap in mass and redshift, except for the 6--7 most massive objects, as discussed previously. Assuming the normalization in B13 is correct, model 2 therefore seems the most plausible method for counting subhalo ancestors. We will have the opportunity to test this model further when results are available for the entire NGVS survey region; at that point the distinct spatial and velocity distributions predicted by the three models (cf.~Figure \ref{fig:model_pv_compare}) should provide an alternative way of selecting between them.

Finally, we also consider the effect of an overall mass rescaling on the SHMR. The mass of Virgo has been argued to lie anywhere in the range 0.33--1.2 times our fiducial value $5.76\times 10^{14}M_\odot$, as discussed in section \ref{sec:simulations}. Since our simulations were rescaled to the fiducial mass, adopting a smaller mass would reduce all subhalo masses by the same factor. On the other hand, adopting a smaller virial radius would increase the fraction of the cluster covered by the pilot survey, and thus the fraction of subhalos located in this region. Adjusting the mass scalings in our simulations, we find these two effects partially cancel; adopting a mass 1.2 times larger has little effect on the pilot region SIMF, while adopting a mass 0.33 times smaller shifts it down by a factor of 2 in subhalo mass. The long-dashed curve in the lower right panel of Figure \ref{Behroozi_comp2} shows the effect of this shift; clearly this would also help reconcile the results in Virgo with the field SHMR determination of B13 at redshifts 0.5--1.5 (thin blue/cyan curves), or the measurement of L12 at $z$ = 0.88 (dashed blue curve).

\section{Conclusion}
\label{sec:conclusion}

Given the complex network of physical processes involved in galaxy formation, simplified descriptions of the final outcome, such as abundance matching, provide an important middle ground between resolved simulations and observational data. In this paper, we have used the stellar mass function for the core of the Virgo cluster recently measured by the NGVS to extend the technique to unprecedentedly low masses. This exercise is fraught with technical challenges. To relate observed galaxies to predicted dark matter substructure, we have to determine how much substructure should survive in the core of the cluster (or more precisely, along a line of sight through the core), trace this substructure back to the point where it last existed as an independent `field' halo, and measure the `infall' mass of that halo at that point. Abundance matching then relates final stellar masses to infall masses, quantifying the net efficiency of star formation in the halos that were to become cluster substructure.

There are uncertainties at each step of this process. Substructure can be disrupted artificially in the simulations by relaxation effects. This problem is relatively easy to correct in an approximate way, by searching through the full merger tree of the cluster to identify structures that should have survived to the present day. On the other hand, this correction introduces several uncertainties into the resulting SIMF. The dense baryon-dominated components of galaxies will help preserve their surrounding dark matter substructure against tidal disruption, for instance, but could also contribute to its disruption through strong stellar winds or similar feedback effects. Thus the true survival rate for substructure is not really known. 

A second problem relates to sub-substructure, that is to say subhalos of subhalos. Physically, these objects could correspond to small satellite galaxies of larger galaxies, or to individual members of galaxy groups that fall into the cluster. In some instances sub-subhalos may merge with their parent subhalos before or while the parents fall into the cluster; in other cases they may be dissociated from their parents and survive as independent objects. Here again, the precise effect on the final SIMF is unclear. 

Finally, we are faced with the usual observational uncertainties -- uncertainties in the global mass and structure of the Virgo cluster, in the completeness of the galaxy detections, particularly at low stellar mass and/or low surface brightness, and in the conversion from light to mass. These problems  {  have already been addressed in Paper I, and will also be discussed further in forthcoming NGVS papers}.

Despite these many caveats, if we combine our most realistic model of the SIMF with our best estimate of the stellar mass function in the core of Virgo, the result is a strikingly simple stellar-to-halo-mass relationship, consistent with previous results such as B13 or L12, but extending these downwards by almost two decades in halo mass or four decades in stellar mass, over the redshift range $z\sim1$--4. Combining our results with those of B13 and L12, we find evidence for a simple broken power-law SHMR over the entire range of stellar mass observed outside the Local Group, from $10^5M_\odot$ or less to almost $10^{12} M_\odot$. This two-part fit to the SHMR in turn suggests {  that} two main mechanisms may set the net efficiency of galaxy formation, one above $\mstar\sim10^{11} M_\odot$, and one below this mass. 

As with most abundance matching studies, our method starts from the assumption that all halos down to some mass limit are occupied by galaxies. An alternative {  possibility} is that galaxy formation is `stochastic', and halos of a given mass can contain a wide range of final galaxy masses. Methods that combine multiple constraints, such as those of B13 or L12, can exclude a large amount of `stochasticity' for luminous galaxies. For dwarf galaxies the situation is less clear, but detailed modeling of the Local Group suggests stochasticity may be important, at least in lower-density environments \citep[e.g.,][and references therein]{Sawala}. If galaxy formation is {  very} stochastic then some halos must be unoccupied, and thus the SHMR for the occupied subhalos must be larger, and must extend to even smaller halo masses, e.g.,~down to $\mhalo\sim 10^9 M_\odot$ rather than $\mhalo\sim 10^{10} M_\odot$. Direct dynamical studies of the smallest Virgo dwarfs may help constrain this possibility, by placing lower mass limits on the dark matter subhalos hosting these galaxies. 

At the lowest stellar masses, we still do not have any examples of galaxy formation outside the Local Group (nor indeed outside the inner halo of the Milky Way). The `ultra-faint' dwarf galaxies \citep{Willman05, Belokurov06, Zucker06}, with stellar masses less than $10^{5} M_\odot$, are currently identified only by star counting, and thus remain invisible to us in more distant systems. It is interesting, however, to consider the implications of our derived SHMR for slightly brighter dwarfs in a system like the Local Group. At a stellar mass of $3\times 10^{5} M_\odot$, {  from our SHMR} we predict an {  average} infall halo mass of $3\times 10^9 M_\odot$. Typically, such a system would be stripped down to two-thirds of this initial {  dark matter} mass, or $2\times 10^9 M_\odot$, by the present day. In the Local Group, we see fewer than 20 satellites per primary above this stellar mass limit \citep{McConnachierev}, whereas from simulations, we expect twice as many subhalos {  over $2\times 10^9 M_\odot$}. Thus, we still have a `missing satellite' problem in the Local Group relative to Virgo: halos that would have been occupied by galaxies, had they been the ancestors of Virgo cluster subhalos, did not host {  visible satellites that survived infall} into the Local Group. There are several possible solutions to this puzzle. The most plausible is that gas cooling and star formation are suppressed in small field halos after reionization \citep{Efstathiou92,BarkanaLoeb,Bullock00,Gnedin06,Sawala}; the difference in subhalo occupation between the Local Group and the Virgo cluster would then relate to the different age distributions of the subhalos' progenitors. We will investigate this possibility further in future work. 

While the results presented here are subject to a number of uncertainties, the prospects for reducing these uncertainties in the near future are excellent. First, our stellar mass function covers only the core of Virgo, less than 1/25th of the total area of the NGVS. Thus we should expect the uncertainties in the stellar mass function to decrease dramatically once the full data set is analysed. The NGVS also has data at other wavelengths, including the near-infrared. We { may} be able to use this to reduce the uncertainties in the stellar mass-to-light ratio { \citep{Courteau14}}. The spatial distribution of galaxies across the whole cluster should help identify which of our models for the SIMF is correct; any variation in the spatial distribution with stellar mass will also allow us to search for redshift variations in the SHMR, as discussed in the Appendix. Currently, redshift information for the cluster is extremely limited relative to the number of galaxies known; in the cluster core, for instance, less than 15\%\ of the galaxies detected in imaging have redshifts in the NASA Extragalactic Database (NED). Ongoing work by the NGVS collaboration and others (see Paper I, section 6) should help to address this, at least for the brighter galaxies. More complete spectroscopy would help confirm which of our estimates of the SIMF is most realistic, and might allow us to detect redshift dependence in the SHMR. In future work we will also complement this additional observational data with higher resolution simulations, so that uncertainties due to resolution effects are reduced. A detailed study of the full galaxy population in Virgo should provide constraints on the efficiency of galaxy formation that are unique and complementary to any others currently available.
 
\acknowledgments

This work was supported by a Post Graduate Scholarship to JG and Discovery Grants to JET and SC, from the Natural Sciences and Engineering Research Council (NSERC) of Canada. JG was also supported by an Ontario Graduate Scholarship from the province of Ontario. EWP acknowledges support from the National Natural Science Foundation of China under Grant No. 11173003, and from the Strategic Priority Research Program, ÓThe Emergence of Cosmological StructuresÓ, of the Chinese Academy of Sciences, Grant No. XDB09000105.
The simulations were performed using the facilities of the Shared Hierarchical Academic Research Computing Network (SHARCNET -- www.sharcnet.ca) and Compute/Calcul Canada. We thank SHARCNET staff for their technical support throughout the project. We also wish to thank the University of Washington HPCC group, Volker Springel, and Alexandre Knebe for making public the codes used to run and analyze the simulations. Finally, we thank Anson Wong, Kyle Oman and Mike Hudson for useful discussions, and the anonymous referee for comments on the initial version of the paper. 

This work is based on observations obtained with MegaPrime/MegaCam, a joint project of CFHT and CEA/DAPNIA, at the Canada-France-Hawaii Telescope (CFHT) which is operated by the National Research Council (NRC) of Canada, the Institut National des Sciences de l'Univers of the Centre National de la Recherche Scientifique (CNRS) of France, and the University of Hawaii. It has been supported in part by the French Agence Nationale de la Recherche (ANR) Grant Programme Blanc VIRAGE (ANR10-BLANC-0506-01), and by the Canadian Advanced Network for Astronomical Research (CANFAR) through funding from CANARIE under the Network-Enabled Platforms program. This research has also made use of the facilities at the Canadian Astronomy Data Centre, which are operated by the National Research Council of Canada with support from the Canadian Space Agency.
 
{\it Facility:} CFHT - Canada-France-Hawaii Telescope

\appendix
\section{Estimating Infall Redshift from Subhalo Orbital Properties}
\label{sec:subhalos}

The main focus of this paper has been abundance matching in the core of the Virgo cluster, using only the total stellar mass function for this region, and ignoring any possible redshift dependence in the SHMR. We expect subhalo properties such as infall redshift to vary systematically with cluster-centric position and line of sight velocity, however. Thus results from the full NGVS survey and/or additional spectroscopic data should allow us to construct multiple samples with different mean $\zinf$, and thus refine our understanding of the correspondence between galaxies and substructure. We explore this possibility further in this appendix.

The correlations between subhalo location, orbital properties and infall redshift have been noted previously in many semi-analytic \citep[e.g.,][]{TB05a} and numerical (e.g., Gao et al. 2011, 2012; Contini et al. 2012 and references therein) studies. Figure \ref{fig:energy} shows the correlation between radial position at $z=0$ (left panel), or infall redshift (right panel), and normalized orbital energy at $z=0$. Subhalos that merge at early times generally occupy lower energy orbits, and the lowest energy orbits are found at smaller radii. The converse is not quite true, however -- there is a population of objects  on weakly bound orbits that are spread over a range of radii (top part of the distribution in the left panel). Further examination shows that these are systems on their first passage in to pericenter, so they are recent mergers.

\begin{figure*}
\epsscale{0.65}
\begin{center}
\includegraphics[scale = 0.5,trim= 0 0 0 170, clip = true]{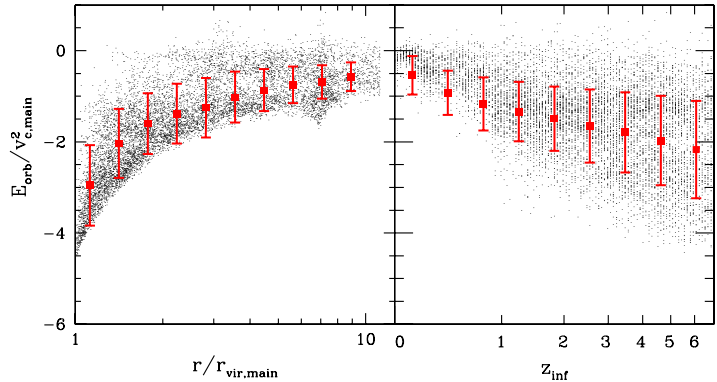}
\caption{Subhalo orbital energy $E_{\rm orb}$, normalized to the square of the main halo's circular velocity at the virial radius, $v_{\rm c}$, versus (3D) radius { $r$ normalized to the virial radius $r_{\rm vir}$} (left) and infall redshift $z_{\rm inf}$ (right). {  For clarity, only one subhalo in 20 is plotted.} Red data points and error bars  indicate the mean and  1-$\sigma$ scatter in evenly spaced bins.\label{fig:energy}}
\end{center}
\end{figure*}

Orbital energy is not directly observable, however; instead we must look for patterns in projected radius $R_{\rm proj}$ and line of sight velocity \vlos. Figure \ref{fig:observables} shows the distribution of subhalos in the space of observable quantities, colored by relative orbital energy (left panel) or infall redshift (right panel). Here we see the separation into recently merged and older material familiar from studies of the phase-space distribution of cluster galaxies { \citep[e.g.,][]{Mahajan,Haines,Kyle,Hou}}. Orbital energy is particularly well correlated with position in phase space. The distribution of infall redshifts is more mixed, but we may still be able to extract useful information from the observables.

\begin{figure*}
\epsscale{0.58}
\plotone{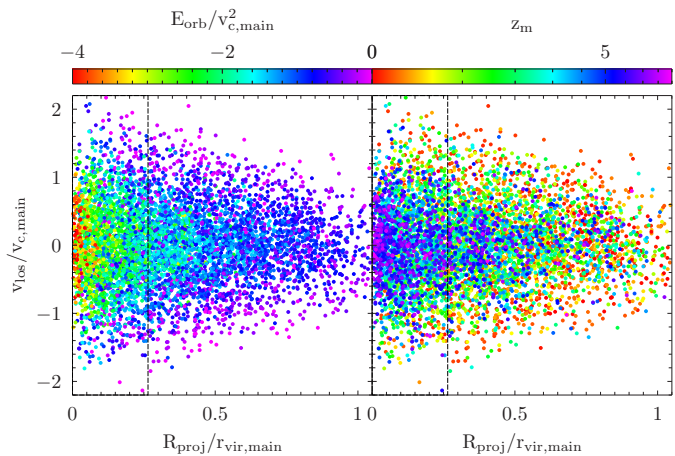} 
\caption{Intrinsic properties in terms of observables. Left: normalized orbital energy {  $E_{\rm orb}/v_{c,main}^2$} as a function of  projected separation $R_{\rm proj}$ and velocity offset $v_{\rm los}$, {  (normalized to the virial radius and mass of the main halo respectively)}. Right: infall redshift $z_{\rm inf}$ in the same phase space.  
 {  For clarity, only one subhalo in 100 is plotted. The vertical dashed line in each panel shows the approximate extent of the pilot region.} \label{fig:observables}}
\end{figure*}

The trend of decreasing $z_{\rm inf}$ with radius is well-known, and has been used for decades to study galaxy transformation in clusters. Data from the central region of NGVS, or similar studies of cluster cores, cannot probe these large-scale radial gradients. There should also be a trend with velocity offset from the cluster center, however, particularly at small projected radii. If we define samples of subhalos by their line of sight velocity {  offset from the mean cluster velocity $v_{\rm los}$}, we can obtain different distributions of mean orbital energy and $z_{\rm inf}$. Figure \ref{fig:velocitybins} illustrates this for cuts at $v_{\rm los}/v_{\rm esc} = 0.2$ and 0.4, where $v_{\rm esc} \equiv \sqrt{2|\phi(x)|}$ is the local escape speed of the cluster. We see that the high-velocity sample in particular contains a large excess of loosely bound systems with $z_{\rm inf} < 0.5$. 

\begin{figure*}
\epsscale{0.56}
\plotone{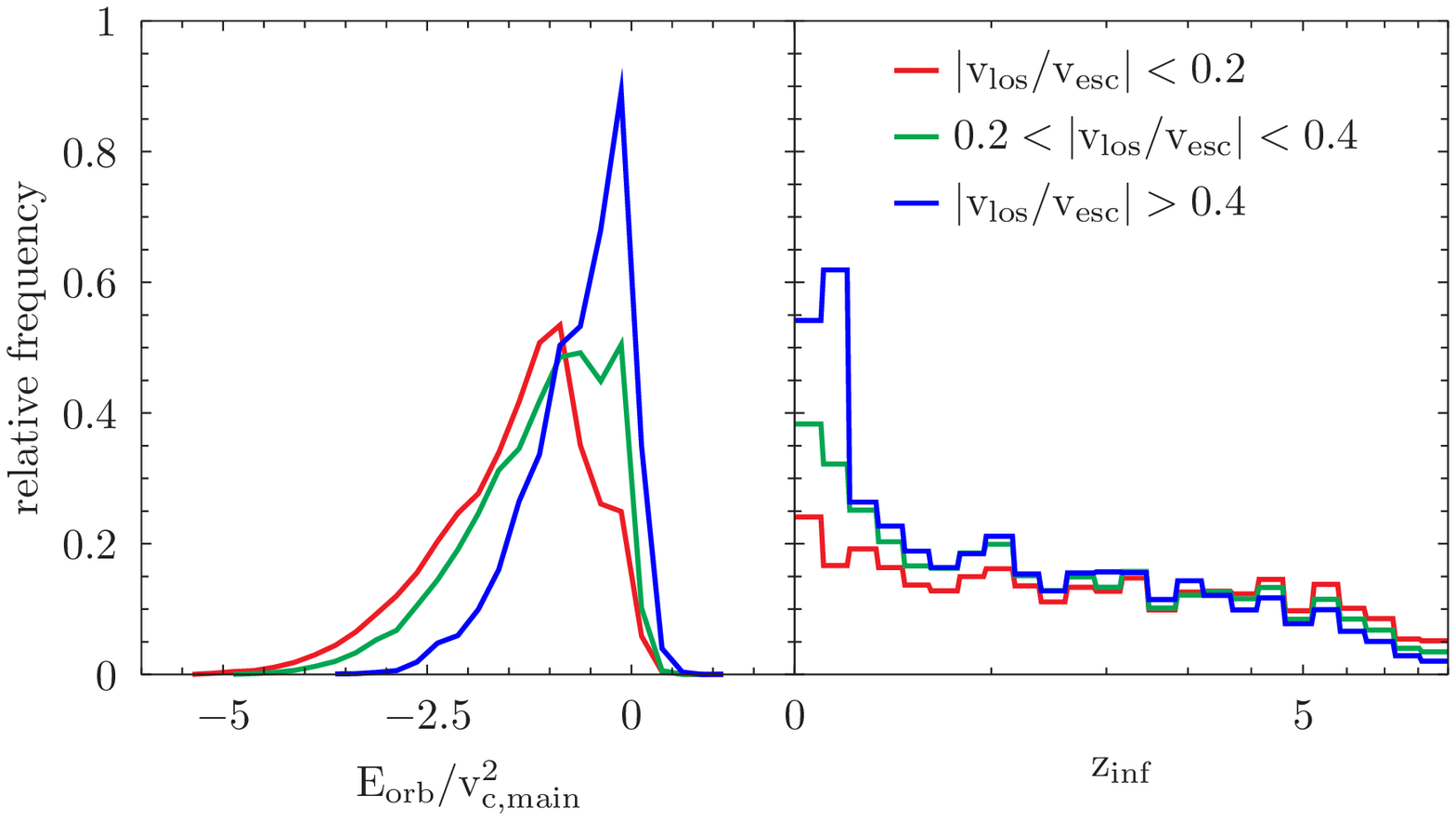}
\caption{Relative distributions of orbital energy and infall redshift for {  all subhalos, binned by normalized line of sight} velocity offset $|v_{\rm los}/v_{\rm esc}|$.}\label{fig:velocitybins}
\end{figure*}

Thus with coverage of the entire {  projected area of the} cluster, and/or more complete spectroscopic information, we will be able to separate Virgo galaxies into subsamples with different mean infall redshifts. This should increase the sensitivity of abundance matching to redshift variations, reduce the uncertainties in the modeling of the SIMF, and constrain stochasticity in subhalo occupation.

\bibliographystyle{apj}
\bibliography{Grossaueretal_rev_aph}

\end{document}